\documentclass[%
 aip,
 amsmath,amssymb,
preprint,%
tightenlines,
]{revtex4-1}

\usepackage{graphicx}
\usepackage{dcolumn}
\usepackage{bm}
\usepackage{subcaption}
\usepackage{color}
\usepackage{upgreek}

\usepackage[utf8]{inputenc}
\usepackage[T1]{fontenc}
\usepackage{mathptmx}

\begin{document}


\title[COVID-19 Urban Bus]{Disease transmission through expiratory aerosols on an urban bus}

\author{Zhihang Zhang}
 \affiliation{%
Department of Naval Architecture and Marine Engineering, University of Michigan
}%
\author{Taehoon Han}
 \affiliation{%
Department of Mechanical Engineering, University of Michigan
}%
\author{Kwang Hee Yoo}
 \affiliation{%
Department of Mechanical Engineering, University of Michigan
}%
\author{Jesse Capecelatro}%
 \altaffiliation[Also at ]{Department of Aerospace Engineering, University of Michigan}
 \affiliation{%
Department of Mechanical Engineering, University of Michigan
}%
\author{Andr\'e Boehman}
 \affiliation{%
Department of Mechanical Engineering, University of Michigan
}%
 
\author{Kevin Maki}
 \email{kjmaki@umich.edu.}
  \affiliation{%
Department of Naval Architecture and Marine Engineering, University of Michigan
}%
%

\date{\today}

\begin{abstract}
Airborne respiratory diseases such as SARS-CoV-2 (COVID-19) pose significant challenges for public transportation. Several recent outbreaks of SARS-CoV-2 indicate the high risk of transmission among passengers on public buses if special precautions are not taken. This study presents a combined experimental and numerical analysis to identify transmission mechanisms on an urban bus and assess strategies to reduce risk. The effects of the ventilation and air-conditioning systems, opening windows and doors, and wearing masks are analyzed. Specific attention is made to the transport of sub-micron and micron-size particles relevant to typical respiratory droplets. High-resolution instrumentation was used to measure size distribution and aerosol response time on a University of Michigan campus bus under these different conditions. Computational fluid dynamics was employed to measure the airflow within the bus and evaluate risk. A risk metric was adopted based on the number of particles exposed to susceptible passengers. The flow that carries these aerosols is predominantly controlled by the ventilation system, which acts to uniformly distribute the aerosol concentration throughout the bus while simultaneously diluting it with fresh air. The opening of doors and windows was found to reduce the concentration by approximately one half, albeit its benefit does not uniformly impact all passengers on the bus due to recirculation of airflow caused by entrainment through windows.  Finally, it was found that well fitted surgical masks, when worn by both infected and susceptible passengers, can nearly eliminate the transmission of the disease. 
\end{abstract}
\maketitle

\section{\label{sec:intro}Introduction}
The SARS-CoV-2 (COVID-19) pandemic has affected people throughout the world, and recovery from the pandemic depends upon a detailed understanding of how transmission occurs in the various ways that humans interact in society.    It is known that among the different pathways of transmission, a dominant mode is that airborne particles carry the virus from person to person~\cite{Zhang20}. To date, transmission of SARS-CoV-2 has predominantly taken place in indoor spaces, especially those with poor ventilation~\cite{world2020ventilation,bhagat2020effects}. It is therefore not surprising that the COVID-19 pandemic poses significant challenges in public transportation. The primary focus of this study is on the factors that contribute to disease transmission on urban buses.

The most documented case of COVID-19 transmission on a bus is from an outbreak on a long-distance coach on January 22 in Hunan, China~\cite{hunan}. Security cameras showed that the contagious individual had not interacted with others on the bus, yet 8 of the 45 passengers were infected over the four-hour journey. Moreover, a passenger was infected who boarded 30 minutes after the contagious passenger disembarked. A similar situation occurred in Zhejiang province around the same time as the Hunan event~\cite{shen2020community}. 128 individuals traveled on two buses to a worship event in Eastern China. It was determined that those who rode the bus with air recirculation enabled had an increased risk of infection compared with those who rode a different bus. It was suggested that airborne transmission may partially explain the increased risk of SARS-CoV-2 infection among the passengers.

Urban buses are an important part of many public transportation systems, and are unique from coach buses in that trips are typically much shorter (tens of minutes), the passengers may be standing or seated, and they make frequent stops.  Although urban buses are heavily used in urban and suburban areas throughout the world, the transmission of airborne particles on urban buses has received little attention. 

The shedding of virus-laden particles is a complicated biological process by which mucus lining the lungs contains the virus, and as air passes through the respiratory tract small droplets are formed and pass through the mouth and into the surrounding air.  The droplets vary in size, from sub-micron to greater than $50~\upmu$m~\cite{bourouiba_violent_2014}. The virus shedding rate is a fundamental quantity that defines the rate at which the virus becomes airborne, yet it is difficult to quantify. The process depends on the individual's breathing rate which varies from person to person, and for an individual depends on activity level, such as resting, walking, speaking, singing, shouting, coughing, sneezing, etc.~\cite{abkarian2020puff}.  The analysis of a superspreading event at a choral rehearsal in the state of Washington in the USA estimates the rate to be around 970~quanta/hr~\cite{Miller20}.  In addition, key factors such as the location of the virus within the respiratory tract and the quantity of virus influence the contagiousness of airborne droplets.   While difficult to directly measure, recent studies~\cite{vuorinen2020modelling} indicate that the shedding rate $\lambda$ is in the range of $1 < \lambda < 50$~s$^{-1}$.

To date, the vast majority of simulation efforts to predict exposure to droplet transmission considers computational fluid dynamics (CFD) where the turbulent air flow is solved using Reynolds-averaged Navier--Stokes (RANS) coupled with Lagrangian particle tracking. Recent examples include transmission of pathogen-laden expiratory droplets (with specific attention on the novel coronavirus; SARS-CoV-2) on buses~\cite{zhu2012,li2017,yang2020transmission,chaudhry2020}, office buildings~\cite{ai2018airborne}, in hospitals~\cite{yu2017ventilation}, and outdoor environments~\cite{feng2020influence}. CFD of aerosol transmission in buses have been studied to assess the influence of filtration ventilation modes, relative humidity, seat arrangement, among other factors in the context of SARS-CoV-2~\cite{yang2020transmission}, influenza~\cite{zhu2012}, and air pollutants~\cite{li2017,chaudhry2020}. Yang et al.~\cite{yang2020transmission} performed numerical simulations to assess the impact of ventilation modes and relative humidity on droplet transmissions in a coach bus. They considered 14 passengers, two size of droplets ($10~\upmu$m and $50~\upmu$m) with five air conditioning supply directions. It was found that ventilation, relative humidity (RH), and initial droplet size significantly influence transmission. It was recommended that high RH, backward supply direction and passengers sitting at nonadjacent seats can effectively reduce infection risk of droplet transmission in buses. Another CFD study analyzed the effects of window openings on self-pollution for a school bus~\cite{li2017}. It was found that opening the driver's window could increase exhaust through window and door gaps in the back of school bus, while opening windows in the middle of the bus could mitigate this phenomenon. Increasing the driving speed was also found to promote higher ventilation rates and further dilute the air. While these studies provide important insight on factors contributing to transmission rates on buses, detailed analyses are limited, especially on urban buses. Even more, experimental measurements of aerosol transmission on buses remain elusive.

The purpose of this study is to investigate the transport of aerosols through an urban bus to identify key factors that contribute to disease transmission and provide guidelines for mitigation strategies.  Experiments are performed to investigate the transport of polydisperse droplets under different settings of the air conditioning system. The experiments also quantify the influence of opening the doors or the windows of the bus. High resolution CFD simulations are performed to determine the transport of small ($<5$ $\upmu$m) particles and investigate the role of the air-conditioning, the location of the infected passenger, the role of face coverings, and the effects of opening the windows and doors. A risk metric is defined based on the number of particles exposed to susceptible passengers.

\section{\label{sec:description}Description of the urban bus, risk of riding, and mitigation strategies}

\begin{figure}[!ht]
    \centering
        \includegraphics[width=0.95\textwidth]{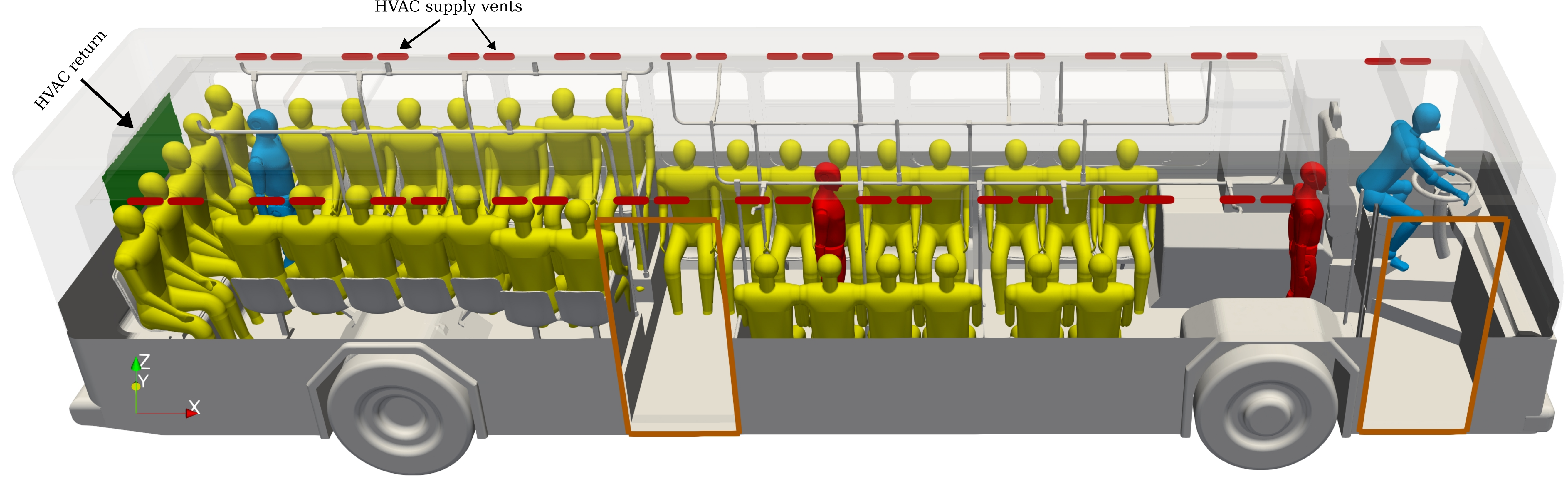}
        \caption{Perspective view of the urban bus interior.}
    \label{fig:BC}
\end{figure}

\begin{figure}[!ht]
    \centering
        \includegraphics[width=0.95\textwidth]{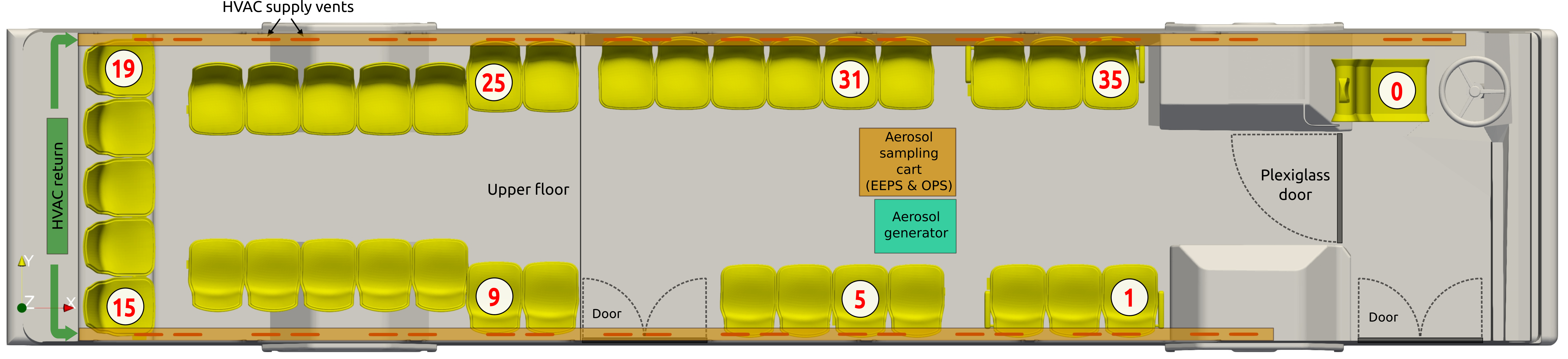}
        \caption{Schematic of the experiment setup.}
    \label{fig:exp-scheme}
\end{figure}

In this work, an urban bus that is used on the campus of the University of Michigan is studied.  The capacity of the bus is 35 passenger seats, with room for up to an additional 30 standing passengers.  The bus makes frequent stops of approximately 30-60 seconds, every one to four minutes.  The longest ride from terminus to terminus of any one of the newly redesigned bus routes is 15 minutes~\cite{siqian}. The bus dimensions are  12.1~m $\times$ 2.58~m $\times$ 2.95~m ($L\times{W}\times{H}$) and a rendering of the bus is shown in Fig.~\ref{fig:BC}.  

The airflow within the bus is affected by the air-conditioning system (heating, ventilation and air conditioning -- HVAC), the opening of windows and doors, breathing, thermal effects, and passenger movement when loading and unloading. The HVAC system can provide a maximum flow rate of 2,500 ft$^3$/min (70.8 m$^3$/min), and the interior volume of the bus is approximately 2,000 ft$^3$ (56.6 m$^3$). The single ventilation  fan draws air from within the passenger compartment through a return vent, and adds 20\% fresh air from outside before returning the air to the passenger compartment through supply vents.  The HVAC return and supply vents are shown in Fig.~\ref{fig:BC}.  The orientation of the HVAC supply vents is such that air exits vertically downward.  The dimension of each supply vent is 9~in by 1~in (0.229~m by 0.0254~m), and the single return vent is 4~ft by 1.5~ft (1.22~m by 0.457~m). A total of 42 supply vents are located along both sides of the bus ceiling, a pair of which are directly above the driver seat. 

There are 14 windows that open, including one near the driver.  The opening part of each window is 10~in tall by 3~ft and 7~in wide (0.25~m by 1.09~m). There are forward and rear loading doors on the passenger side of the bus.  A transparent shield door is installed between the driver and passenger area to impede virus transmission between the two areas so that only the rear door is used for loading and unloading.

Much of the work towards mitigation in public spaces is based on the distance that should be kept between people, commonly referred to as social distancing.  The early work~\cite{Wells34} demonstrates that the larger heavier particles, those greater than 100~$\upmu$m, fall within 2~m of being exhaled. This principle is used throughout the world for socially distancing guidelines, but it does not account for the influence of convection of the small particles that travel with the ambient air currents. Furthermore, it is becoming clear that very small particles, those which do not fall to the ground, stay suspended in air, and travel passively with the ambient air flow~\cite{Zhang20}.  In order to safely use urban buses it is important to understand virus transport via the smallest particles so that effective mitigation strategies can be implemented.  Towards this goal in this paper high resolution numerical simulations are conducted to  predict the travel of particles that are subjected to all of the relevant forces that govern its transport through the passenger cabin, with particular attention to the turbulent flow that dominates the transport of the small aerosols.

The risk associated with riding a bus is quantified by directly calculating the number of inhaled particles at each location on the bus with contagious passengers located in different positions. Additionally  the role of masks is demonstrated by using a simple model of mask effectiveness based on recent literature~\cite{Dbouk20}.  Finally, the influence of using the variable speed HVAC system, and opening the windows and doors  is quantified.

The infected passenger is characterized as shedding the virus at the highest suggested rate~\cite{vuorinen2020modelling} of 50~s$^{-1}$. This number is based on literature and analysis of several spreading events in Asia and Europe. The shedding rate represents a worst-case scenario, corresponding to a highly contagious passenger speaking loudly and continuously throughout the bus ride.  It is assumed that only one infected passenger is present, and the analysis investigates transmission with the infected passenger either standing in the front or in the middle of the bus.

\section{\label{sec:exp}Experimental analysis}

\subsection{Experimental setup}

One aerosol generator and two sampling instruments were utilized to measure aerosol transport and dispersion and emulate an infected passenger on an urban bus. A theatrical fog machine (CO-Z Portable Fog Machine - 400 watt) was used to generate aerosols using a water-based `fog-juice.' The nontoxic water-based fog-juice is comprised of deionized water, propylene glycol (C$_{3}$H$_{8}$O$_{2}$, CAS number 57-55-6), and triethylene glycol (C$_{6}$H$_{14}$O$_{4}$, CAS number 112-27-6). The injection time of all cases was three seconds to generate sufficient mass and consistent concentration of aerosol. 

\begin{figure}[!ht]
    \centering
        \includegraphics[width=0.65\textwidth]{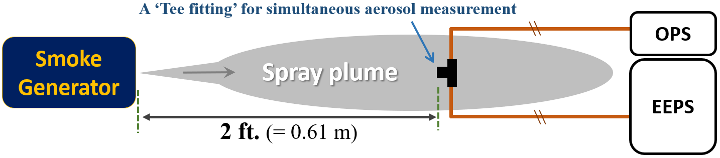}\\
        \includegraphics[width=0.65\textwidth]{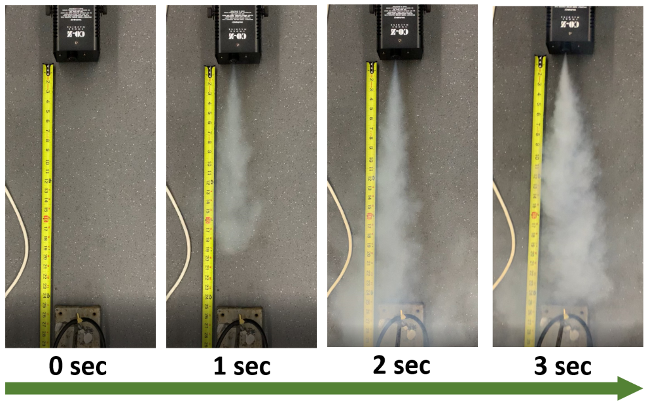}
        \caption{Schematic and pictures of sampling configuration for evaluation of size distribution from the smoke generator}
    \label{fig:exp-pics}
\end{figure}

The target range of aerosol size measurement was from 5~nm to 10~$\upmu$m (10,000~nm) to include the size of the virus itself, virus-containing aerosols, and droplets. Two different types of instruments were used in this study: (i) a TSI EEPS (Engine Exhaust Particle Sizer) Model 3090 for measuring nano-sized aerosol size and numbers, and (ii) a TSI OPS (Optical Particle Sizer) Model 3330 for quantifying micro-sized aerosol size and numbers. The two instruments were connected at a unified sampling location via a tee fitting, as shown in the upper schematic diagram in Fig.~\ref{fig:exp-pics}. 

The EEPS instrument measures the number and size distribution of aerosols from 6 to 520~nm range with high temporal resolution (10 Hz), which permits instant visualization of aerosol dynamics during transient events. The TSI Optical Particle Sizer (OPS) Model 3330 is a portable instrument that also provides a fast measurement of aerosol concentration and size distribution using an optical particle counting technology. The OPS instrument has a size range from 0.3~$\upmu$m to 10~$\upmu$m with 1~Hz time resolution. 

The two instruments were mounted in a stacked configuration on a cart to permit movement to different sampling locations on the bus. Thus, the release of aerosol in different locations of the bus was enabled by the use of the portable smoke generator and the sampling at different locations in the bus enabled assessment of aerosol transport times and aerosol dilution throughout the bus. The instruments were benchmarked each time using ambient aerosols and a HEPA filter (99.97\% capture for particles larger than 3~$\upmu$m) for accurate measurements.

Both of the aerosol measuring instruments count the particle numbers in a specific range of aerosol diameter. The EEPS measures 22 electrometer channels and draws 32 aerosol diameter sizes from 6~nm to~520 nm, while the OPS measures 16 ranges of aerosol diameters from 0.3~$\upmu$m to 10~$\upmu$m. A conversion equation is necessary to calculate the total concentration for each instrument for each bin that comprises a particle size range. The calculation method for total concentration (total number) used for the EEPS and OPS data processing is given by
\begin{equation}
\label{eqn:exp}
N = \int_{D_{p1}}^{D_{p2}} \frac{dN}{d\log{D_p}}d\log{D_p},
\end{equation}
where $D_p$ is the channel midpoint of particle diameter, and $N$ represents the concentration in a specific range of diameter ($D_p$). $D_{p1}$ and $D_{p2}$ are the target range of the aerosol total concentration. In this study, the EEPS used the $D_{p1}$ and $D_{p2}$ values as 6 and 520~nm, and the OPS used 300~nm to 10~$\upmu$m for containing the maximum range of the instrumental diameter size windows. The total concentration is expressed as a concentration size spectral density in $dN/d\log{D_p}$ (cm$^{-3}$) with units of $N$ (cm$^{-3}$). The logarithmic term arises from the fact that the size classes are logarithmically spaced. In order to convert $dN/d\log{D_p}$ (cm$^{-3}$) to $N$(cm$^{-3}$), the $dN/d\log{D_p}$ values of interest were summed and divided by the number of channels for each instrumental value. 

\begin{figure}[!ht]
    \centering
        \includegraphics[width=0.45\textwidth]{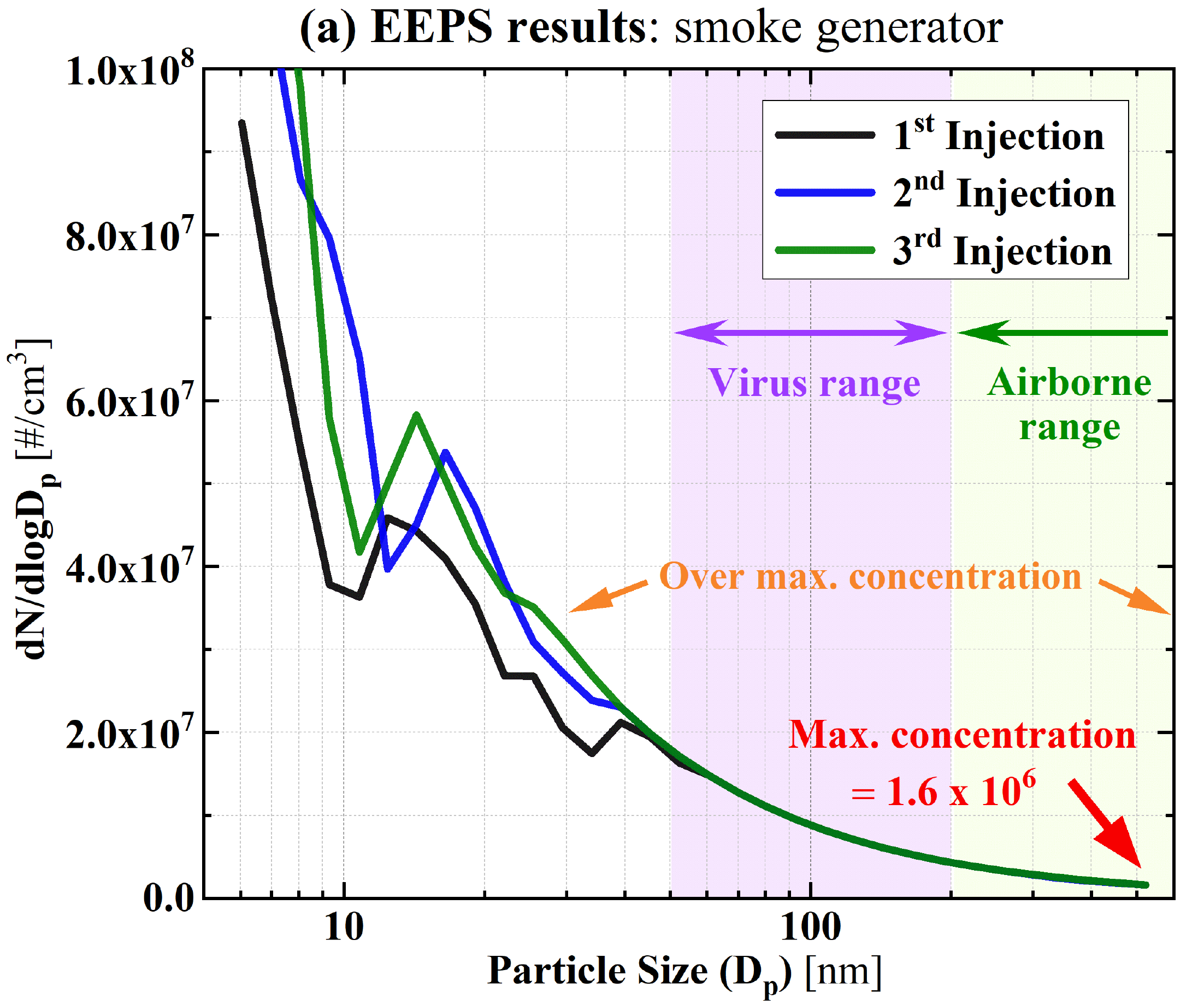}
        \includegraphics[width=0.45\textwidth]{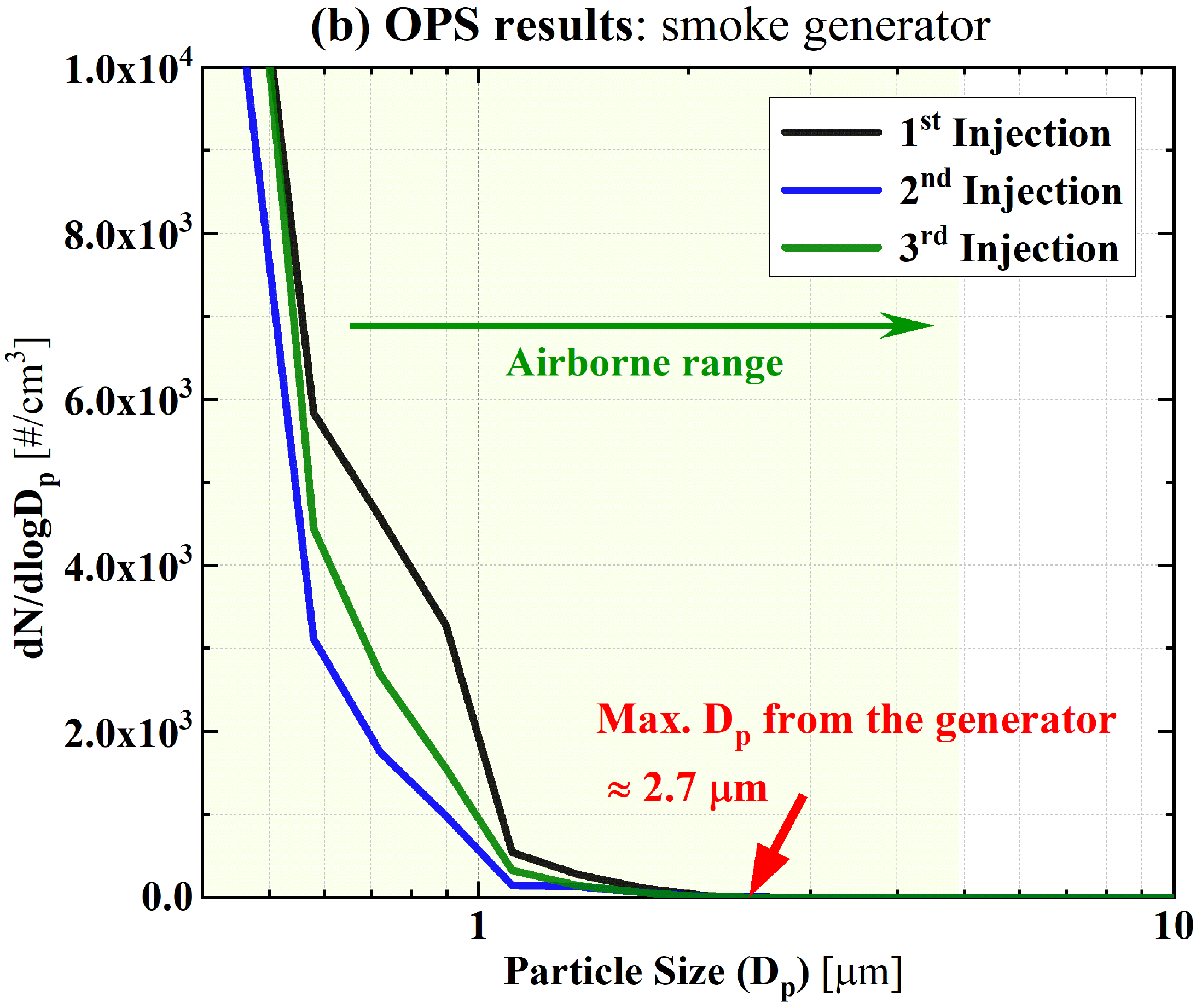}
        \caption{Smoke generator emitting aerosol size distribution and concentrations: (a) EEPS -- nano range, (b) OPS -- micro range.}
    \label{fig:eeps-ops}
\end{figure}

Prior to conducting the experiments, the aerosols emitted from the smoke generator were measured using the two instruments. Fig.~\ref{fig:exp-pics} illustrates a schematic of  aerosol generation and measurement setup (top figure), and images of the smoke plume at different instances in time (bottom pictures). Fig.~\ref{fig:eeps-ops} shows the results of the aerosol size distribution and concentrations from the smoke generator spray plume (including three repeated measurements). Fig.~\ref{fig:eeps-ops}(a) is the nano range aerosol EEPS result, and Fig.~\ref{fig:eeps-ops}(b) is the micro range aerosol OPS result. The particle concentration over 30~nm diameter shows consistent results for all three experiments, and it is due to the exceeded maximum concentration limit of the EEPS instrument. The concentration of small size aerosol is over $10^8$, and the largest aerosol diameter limit is over $10^6$, the smoke generator emits sufficient mass and size distribution of aerosol for these experiments. The OPS results in Fig.~\ref{fig:eeps-ops}(b) also showed a maximum concentration as particle size approaches 500 nm diameter, and the smoke generator emitted a maximum particle diameter around 2.7~$\upmu$m as indicated by the arrow. Based on the literature the virus size is around 50 to 200 nm~\cite{Yang2020covid, Chen2020epidemiological} (purple shadowed area in Fig.~\ref{fig:eeps-ops}(a)) and the virus carrying aerosol size is up to 5~$\upmu$m(green shadowed area in Fig.~\ref{fig:eeps-ops}(a) and (b)). Thus the aerosols are sufficient for representing the target aerosols in this study.

\subsection{Experimental results}

Measurements were taken to assess the influence of location on bus and the effects of having the windows open or closed. Each condition is classified by the locations of aerosol sampling and injection.  The sample location is denoted as A, B, or C, corresponding to the driver (seat 0), front passenger (seat 31), or middle passenger (seat~9), as depicted in Fig.~\ref{fig:exp-scheme}.  The three injection locations are denoted  1, 2 or 3, corresponding to the front (seat 5), middle (seat 9) or back (seat 15).  Table~\ref{tab:exp} summarizes the experimental measurement locations.

\begin{table}
\centering
\caption{Experimental measurement locations \label{tab:exp}}
\begin{ruledtabular}  
\begin{tabular}{cccc}
\bf \# & \bf Case & \bf Sampling Seat & \bf Injection Seat \\\hline
\bf 1&	Ambient&	\# 0 (Driver)&	none\\\hline
\bf 2&	A - 1&\# 0 (Driver)&	\# 5 (Front)\\
\bf 3&	A - 2&\# 0 (Driver)&	\# 9 (Middle)\\
\bf 4&	A - 3&	\# 0 (Driver)&	\# 15 (Back)\\\hline
\bf 5&	B - 1&	\# 31 (Front)&	\# 5 (Front)\\
\bf 6&	B - 2&	\# 31 (Front)&	\# 9 (Middle)\\
\bf 7&	B - 3&	\# 31 (Front)&	\# 15 (Back)\\\hline
\bf 8&	C - 1&	\# 9 (Middle)&	\# 5 (Front)\\
\bf 9&	C - 2&	\# 9 (Middle)&	\# 9 (Middle)\\
\bf 10&	C - 3&	\# 9 (Middle)&	\# 15 (Back)\\
\end{tabular}
\end{ruledtabular}
\end{table}

Table~\ref{tab:expLocations} shows the detailed geometry of sampling and injection points, and the measured distance ($D$) and height ($h$) values are from the sidewall and the floor on the bus. The direction of aerosol injection and sampling followed the passenger and driver face direction of the seats, so the front (1, B) and middle (2, C) seats are toward bus central direction and the driver (A) and back (3) seats are front direction as shown in Fig.~\ref{fig:exp-scheme}. 

\begin{table}
\centering
\caption{Sampling and injection location ($D$: distance from the sidewall, $h$: height from the floor). \label{tab:expLocations}}
\begin{ruledtabular}  
\begin{tabular}{ccccc}
\bf Type& \bf Location&	\bf Seat \#& \bf Distance ($\bf{D}$)& \bf Height ($\bf{h}$)\\\hline\
Sampling (A)&	Driver	&\# 0	&21.0 in (0.53 m)&	35.0 in (0.89 m)\\
Sampling (B)	&Passenger - Front	&\# 31   &31.0 in (0.79 m)  &42.0 in (1.07 m)\\
Sampling (C)&	Passenger - Middle	&\# 9	&31.0 in (0.79 m)	&45.0 in (1.14 m)\\\hline
Injection (1)&	Passenger - Front	&\# 5	&22.0 in (0.56 m)	&20.5 in (0.52 m)\\
Injection (2)&	Passenger - Middle	&\# 9	&25.0 in (0.64 m)	&21.5 in (0.55 m)\\
Injection (3)&	Passenger - Back	&\# 15   &23.5 in (0.60 m)	&23.5 in (0.60 m)\\
\end{tabular}
\end{ruledtabular}
\end{table}

The baseline experiments were conducted in a stationary bus with the windows and doors closed. The aerosol numbers and concentration are sensitive to the ambient environment (such as temperature, pressure, and humidity) so each case of experiments was repeated at least two times for all conditions and on different days for the baseline experiments. In this study, the aerosol response time was calculated as the time difference between the aerosol injection and the initial slope change of the total concentration in all cases.

\begin{figure}[!ht]
    \centering
        \includegraphics[width=0.45\textwidth]{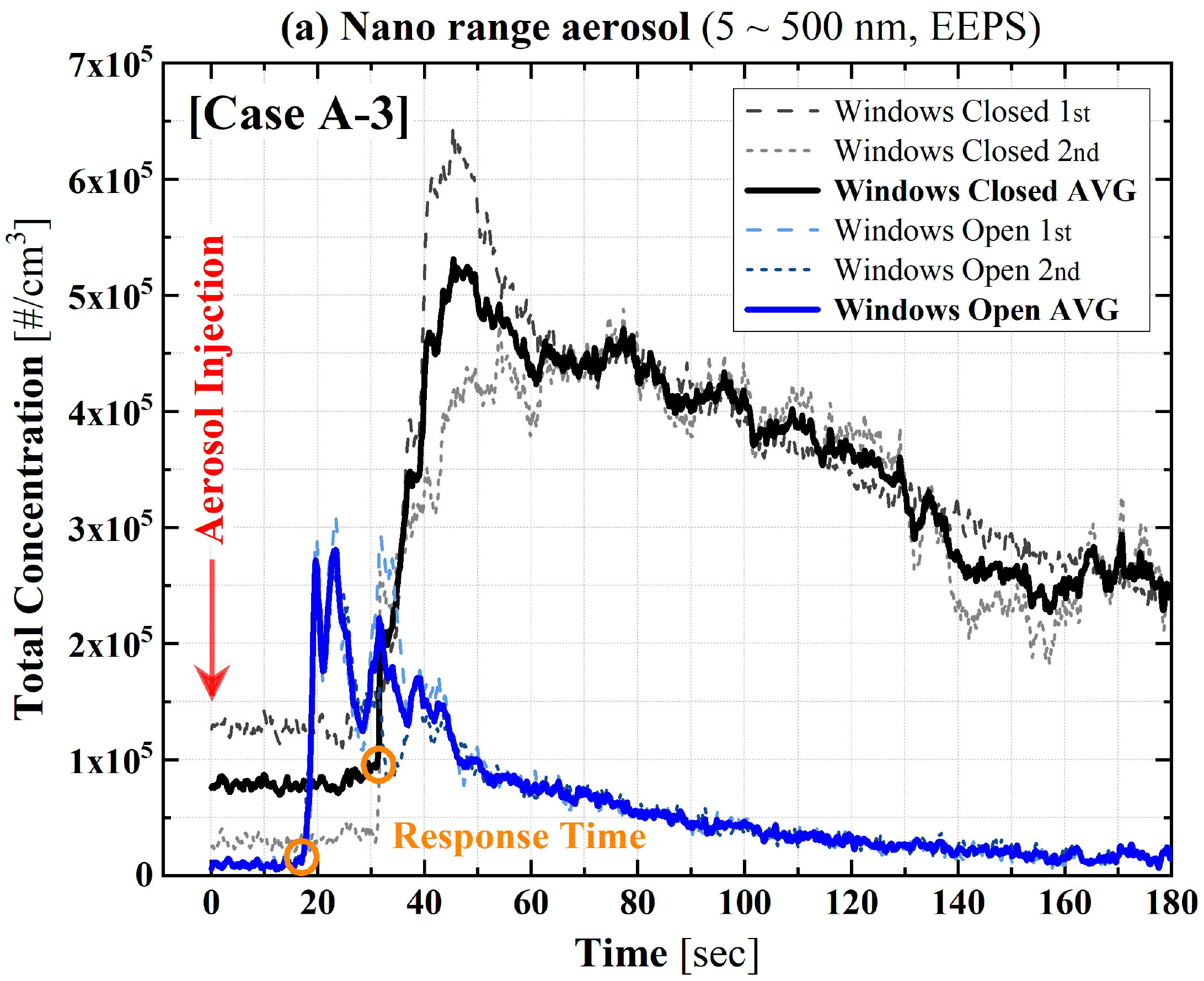}
        \includegraphics[width=0.45\textwidth]{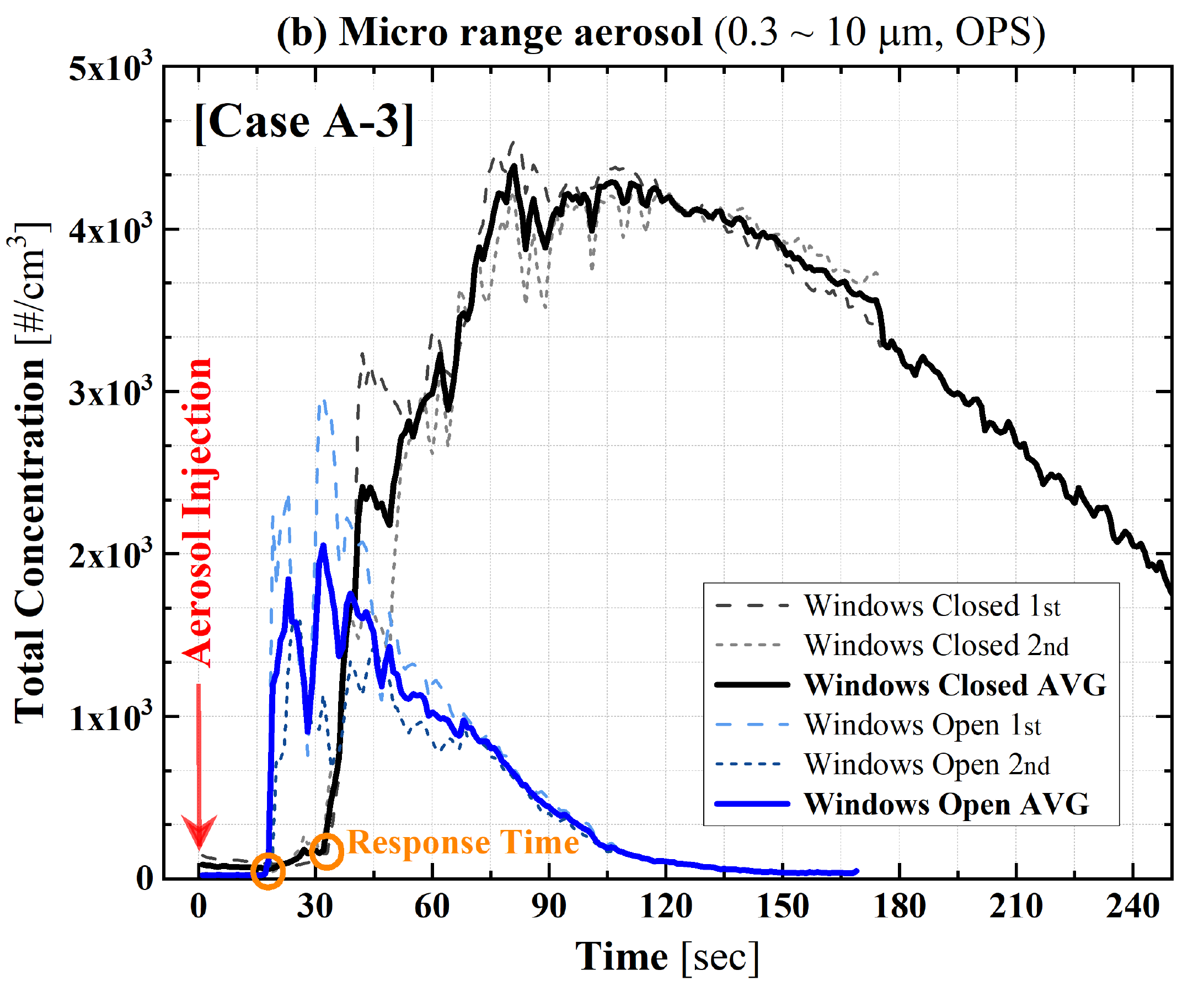}
        \caption{Total concentrations with and without windows open: (a) nano-sized aerosols, (b) micro-sized aerosols. Case A-3.}
    \label{fig:eeps-ops-open-closed}
\end{figure}

Fig.~\ref{fig:eeps-ops-open-closed} shows the time history of the aerosol concentration measured with both instruments for Case A-3 when the bus is stopped but the windows are either open or closed.  The measurements are repeated two times, and the average is shown in the dark line. The first observation is that the effect of opening the windows is significant to reduce the  concentration. The  concentration of nano range aerosol in Fig.~\ref{fig:eeps-ops-open-closed}(a) is reduced by half with windows open, and the micro size aeroslol has the same trend as shown in Fig.~\ref{fig:eeps-ops-open-closed}(b). Also, it is remarkable to see that the aerosol response time is greatly shortened with windows open for both size range aerosols. The reduced response time could be due to the promoted air mixing with windows open.

\begin{figure}[!ht]
    \centering
        \includegraphics[width=0.3\textwidth]{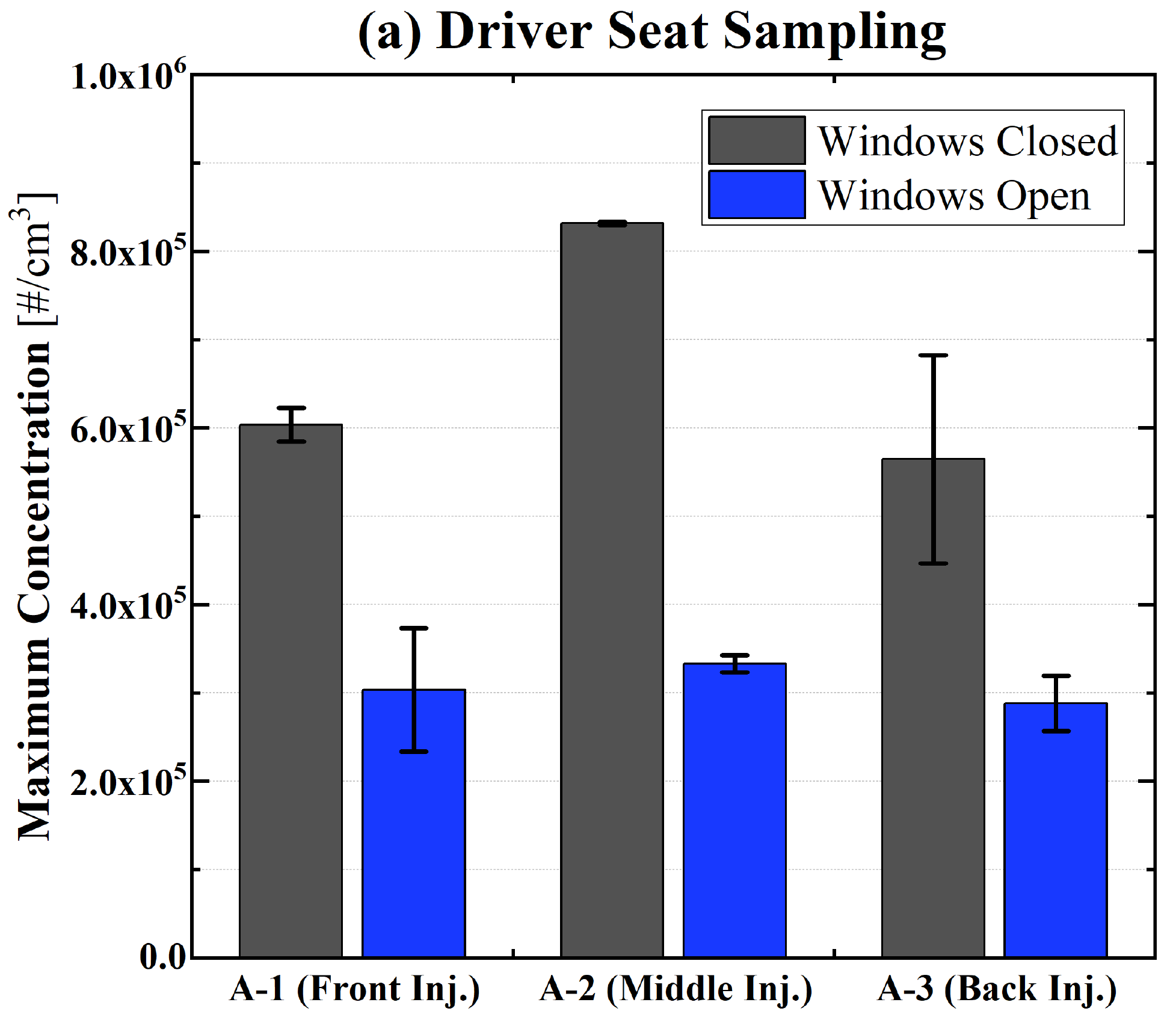}
        \includegraphics[width=0.3\textwidth]{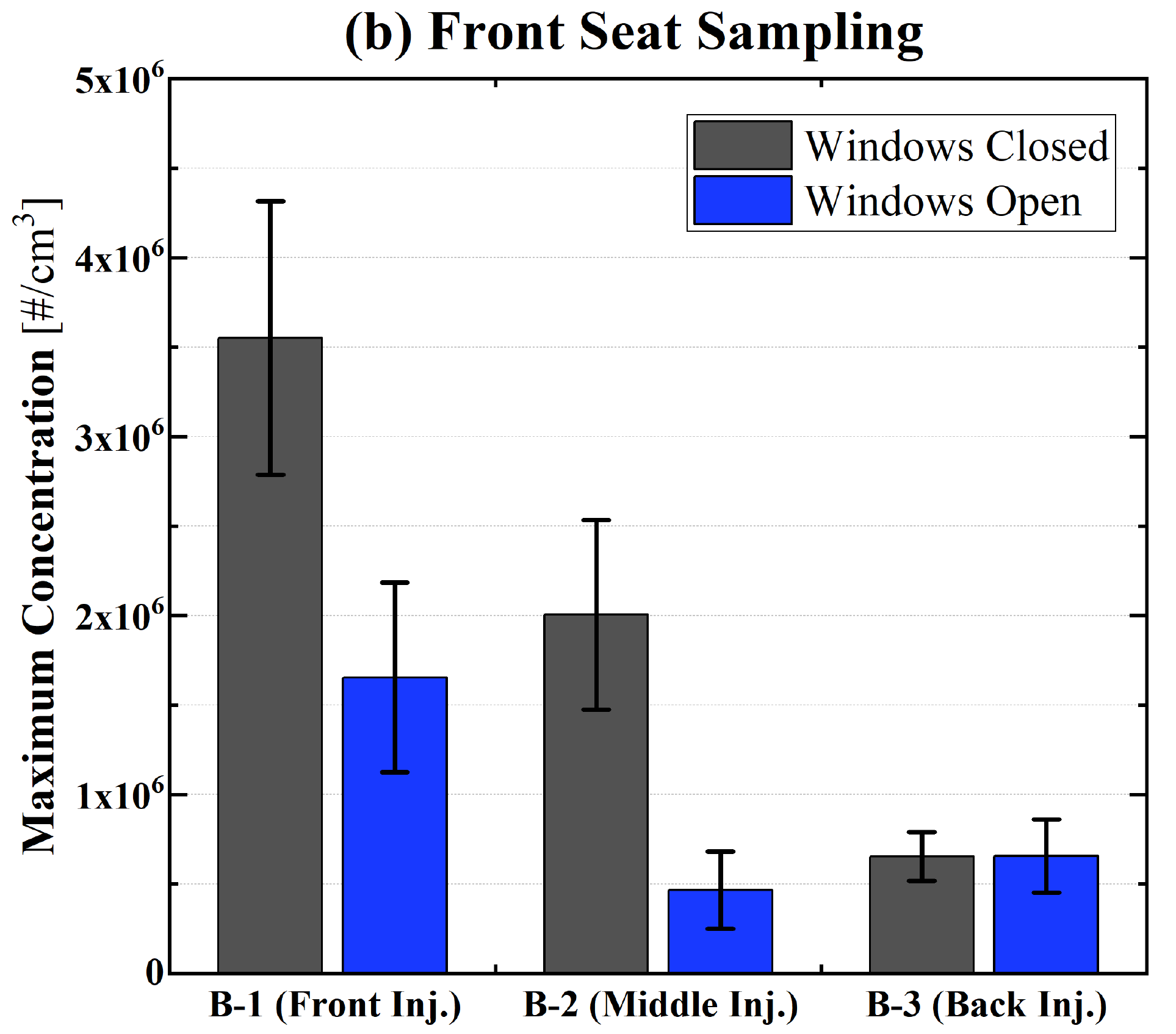}
        \includegraphics[width=0.3\textwidth]{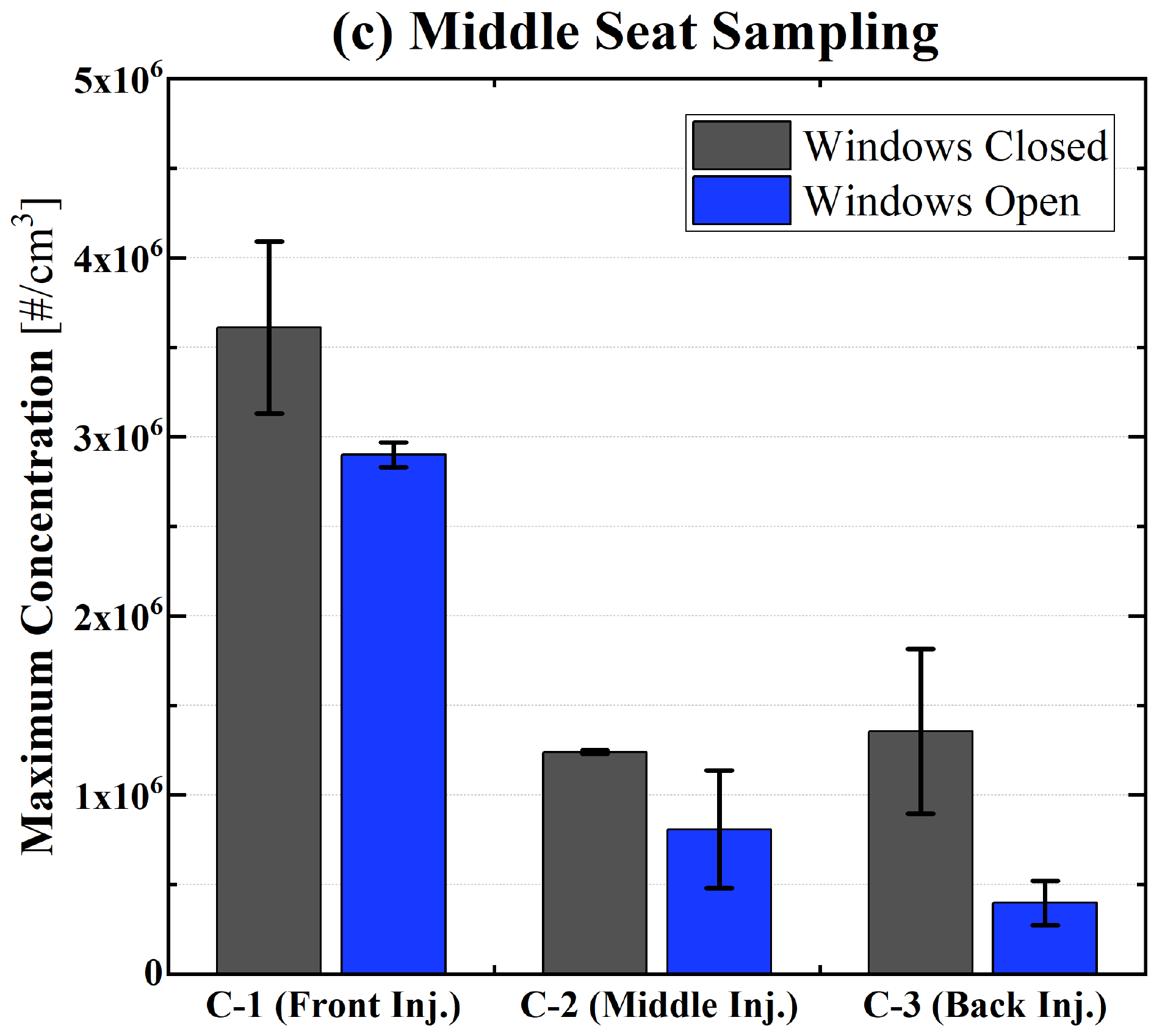}

        \caption{Nano-sized aerosol maximum concentration comparisons for windows open and closed conditions: sampling at (a) Driver seat, (b) Front seat, and (c) Middle seat.}
    \label{fig:open-close-summary}
\end{figure}

Fig.~\ref{fig:open-close-summary} depicts the summary of all 9 cases of experiments. Similar to Case A-3 in Fig.~\ref{fig:eeps-ops-open-closed}, the maximum concentration is reduced by approximately 50~\% when the windows are opened for all cases.

\begin{figure}[!ht]
    \centering
        \includegraphics[width=0.3\textwidth]{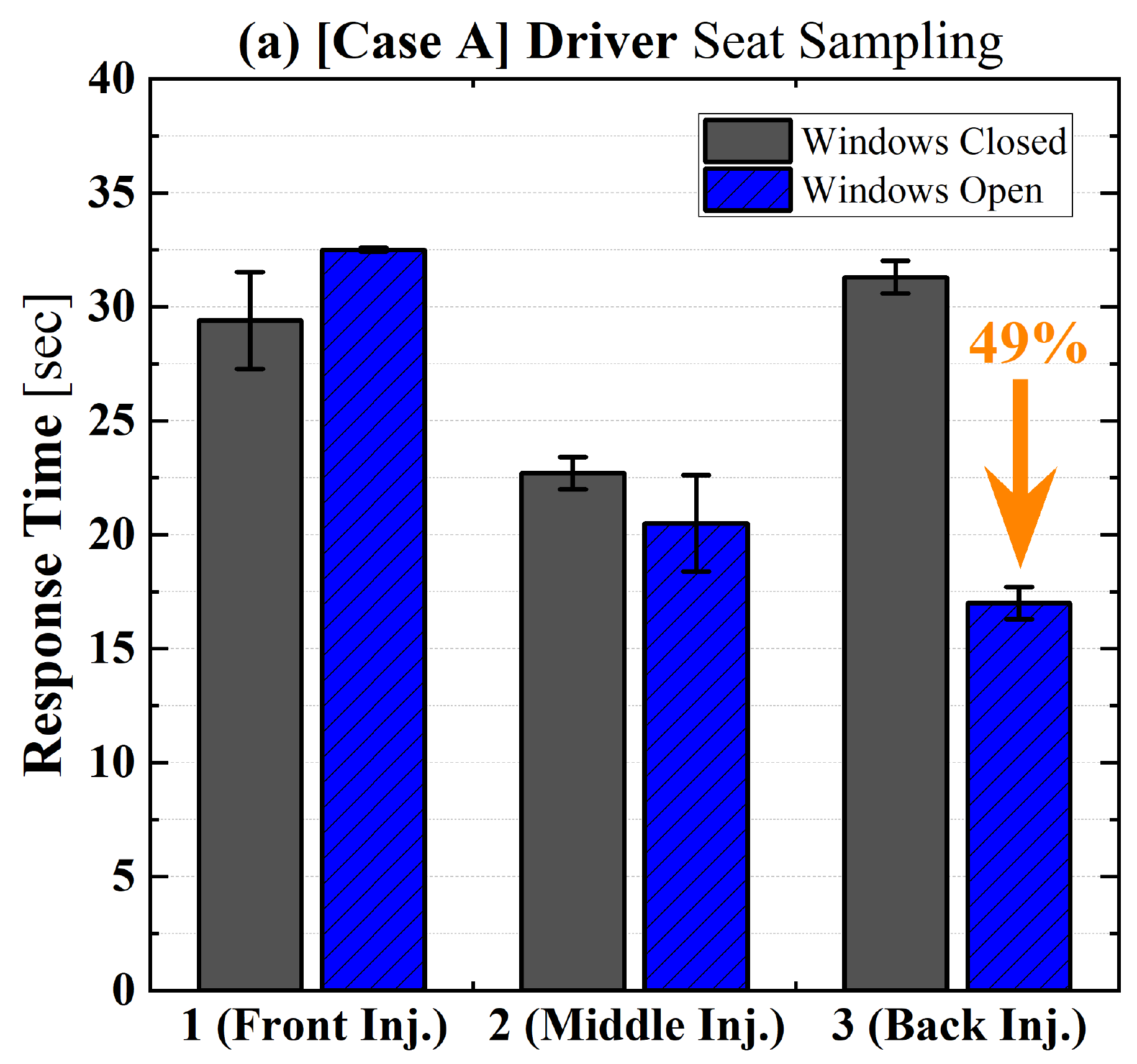}
        \includegraphics[width=0.3\textwidth]{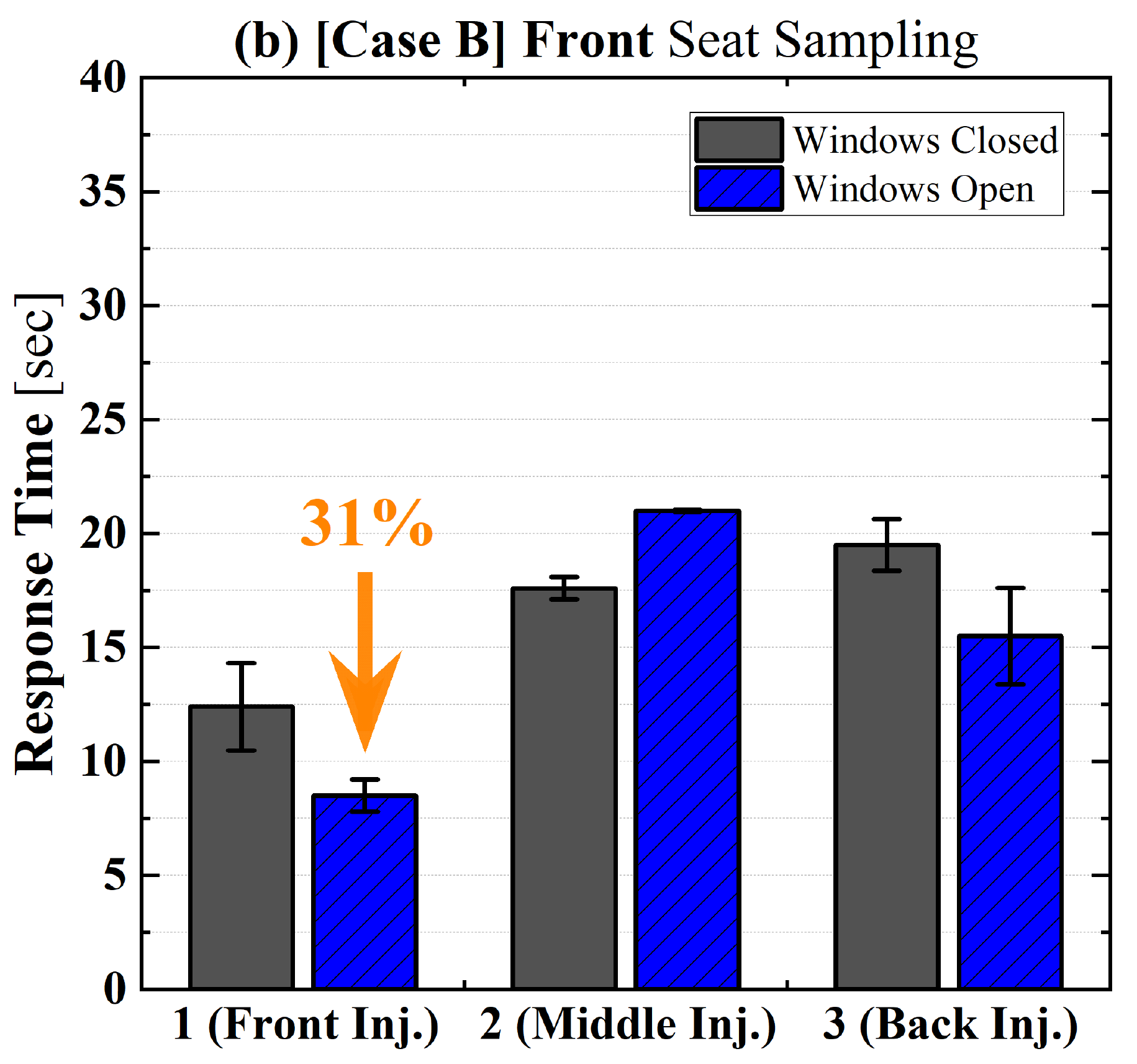}
        \includegraphics[width=0.3\textwidth]{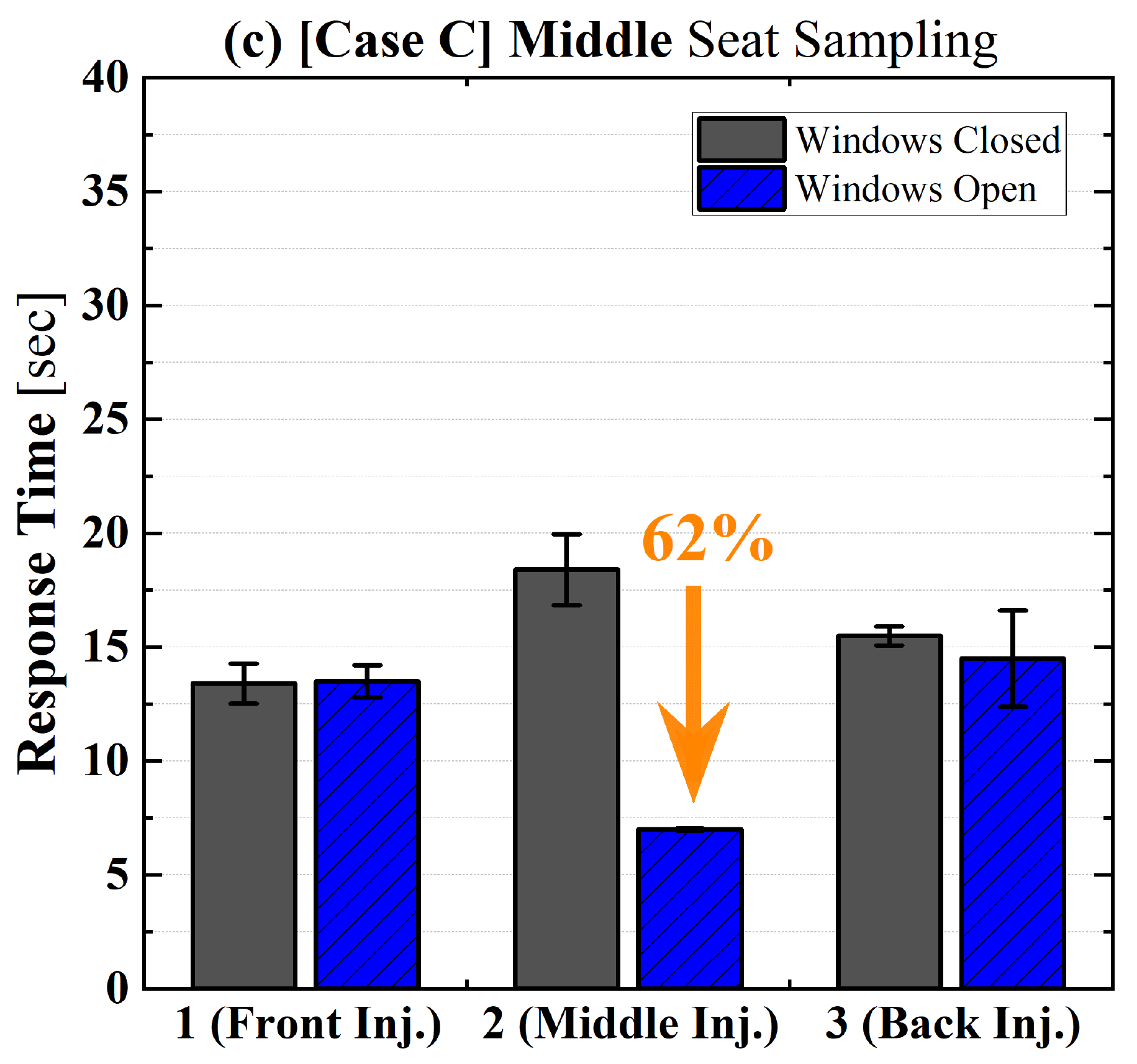}

        \caption{Response time comparison between windows closed and open conditions: sampling at (a) Driver seat, (b) Front seat, and (c) Middle seat.}
    \label{fig:open-close-response}
\end{figure}

Fig.~\ref{fig:open-close-response} shows the summary of response time with the windows open and closed. Although there is no significant variations between the windows open and closed conditions, when the distance between the injection sampling locations are short, the response time is reduced with windows open.
 
\section{Computational analysis}

The Reynolds-Averaged Navier-Stokes (RANS) equations are numerically solved to predict the turbulent flow field inside and around the bus.  An energy equation is used to account for the influence of temperature variations. The virus-laden aerosols are modeled as a continuum  in which the concentration density evolves according to its transport equation. The RANS equations together with the energy and aerosol concentration equations are solved using a customized solver based on the OpenFOAM opensource CFD library.

\subsection{Numerical solver and governing equations}

\subsubsection{Modeling of the airflow}

The flow inside and around the bus is assumed to be incompressible and turbulent.  The unsteady  RANS equations represent conservation of mass and momentum, and are expressed as
\begin{eqnarray}
\nabla\cdot\mathbf{u}&=&0\\
\frac{\partial{\mathbf{u}}}{\partial{t}}+\nabla\cdot(\mathbf{u}\mathbf{u})&=&-\nabla{p_{\rm rgh}}-\mathbf{g}\cdot\mathbf{x}\nabla\left(\frac{\rho}{\rho_{0}}\right)+\nabla\cdot\left[\nu_{\rm eff}(\nabla\mathbf{u}+\nabla\mathbf{u}^{T})\right],
\end{eqnarray}
where $\mathbf{u}$ and $p_{\rm rgh}$ are the Reynolds-averaged velocity vector and kinematic pressure, $\mathbf{g}$ is the acceleration due to gravity, and $\nu_{\rm eff}$ is the effective viscosity that accounts for both molecular and turbulent diffusion. The Boussinesq approximation is used in this model such that the density difference is ignored except in the gravitational term~\cite{ferziger2002computational}. The nominal air density is $\rho_0$, and the local density $\rho$. 

The kinematic pressure represents the difference between the total pressure and the hydrostatic pressure, i.e. $p_{\rm rgh}=(p-\rho\mathbf{g}\cdot\mathbf{x})/\rho_{0}$. The local density is determined from the local temperature according to $\rho/\rho_{0}=1-\beta(T-T_{0})$. Here, $\beta$  is the thermal expansion coefficient and takes a value of $\beta = 3\times10^{-3}$, and $T$ and $T_{0}$ are the local and nominal temperatures, respectively.

The temperature variations between the outside air, the cooled air coming from the air-conditioning system, and the passengers can generate flows due to buoyancy effects.  As the temperature varies, so does the air density. For small variations in temperature (approximately 10~$^{\circ}$C), and flow speeds much less than the speed of sound, the density variation can be neglected in the continuity equation and momentum equation with the exception of the gravity term~\cite[pp 117-121]{Kundu08-FluidMechanics}.

The equation governing temperature is

\begin{equation}
\frac{\partial{T}}{\partial{t}}+\nabla\cdot(\mathbf{u}T) - \nabla\cdot({\alpha_{\rm eff}}\nabla{T}) = 0,
\end{equation}
where $\alpha_{\rm eff}=\nu_{t}/{\rm Pr}_{t}+\nu/{\rm Pr}$, and ${\rm Pr}_{t} = 0.9$ and ${\rm Pr} = 0.71$ are the turbulent and laminar Prandtl numbers, respectively. The $k-\epsilon$  turbulence model is used to determine the turbulent viscosity~\cite{launder1983}. 

\subsubsection{Modeling of aerosol transport}

When humans breathe, cough, sneeze, sing, etc., small droplets are exhaled into the surrounding air.  For normal conditions such as breathing, speaking, and even coughing, most of the exhaled droplets are under $1~\upmu$m in diameter, and rarely larger than $5~\upmu$m~\cite{papineni1997size,fabian2008influenza}. This has significant relevance on the distance that an exhaled particle can travel. While larger particles are dominated by gravity and are pulled vertically downwards, the smallest particles are  neutrally buoyant, and move passively with the carrier fluid. To evaluate the role of gravity on the particle trajectory, the Stokes number of the droplets can be calculated as ${\rm St}=\tau_{p}/\tau_{f}$, where the droplet kinematic timescale $\tau_{p} = \rho_{p}d_{p}^2/(18\nu\rho_{f})$, with $\rho_{p}$ and $\rho_{f}$ the densities of the particle and the fluid, respectively. For the majority of the exhaled droplets ($d_{p}<1~\upmu$m), the droplet kinematic timescale $\tau_{p}<1$~$\upmu$s. In an indoor environment, the fluid timescale $\tau_{f}$ is larger than 1~s.  The present numerical simulations focus on the transport of the droplets that travel passively with the carrier fluid, which are those with a diameter less than $5~\upmu$m. 

The exhaled particles are described as an aerosol concentration of droplets per unit volume, $C(\mathbf{x},t)$. The concentration field $C$ is governed by the convection-diffusion equation:
\begin{equation} 
\frac{\partial{C}}{\partial{t}}+\nabla\cdot(\mathbf{u}C)-\nabla\cdot\left(\alpha_{\rm eff}\nabla{C}\right)=0.
\end{equation}

The equations governing the fluid flow, temperature, and aerosol concentration are solved using the OpenFOAM opensource CFD library.  All discretization schemes are nominally second-order in space and time.

\subsection{Computational setup and case designs}

The bus geometry, including the interior of the cabin, windows, doors, seats, hand rails, ventilation supply and return (see Fig.~\ref{fig:BC}), are determined from a laser scanner and used for generating the computational grid of the fluid domain. Manikins are placed at different locations inside the bus: a driver sitting behind the wheel and standing passengers. 

The infected passenger has a shedding rate of 50~s$^{-1}$ and a continuous breathing assumption is used so that the velocity on the mouth of the infected passenger is always outward at a  breathing rate of 0.1 l/s. A turbulence intensity of 2.5$\%$ and a turbulence length scale of $5\times10^{-3}$~m are enforced at the supply vents for cooled air at 20~$^{\circ}$C. For the breath of the passengers, the turbulence intensity and tubulence length scale are 10$\%$ and $7.5\times10^{-3}$ m. A normal oral temperature of 37~$^{\circ}$C is applied for the mouths of passengers as boundary conditions. The remaining surfaces  of the cabin are assumed to be no-slip and adiabatic walls.

A precursor simulation of three minute duration is performed to generate a fully developed turbulent flow field inside the passenger cabin. The resulting flow field is used for the initial condition for the simulation and analysis of the aerosol transport.  

A series of simulations are performed to assess the numerical uncertainty, the influence of the HVAC system, the influence of the location of the infected passenger, and the roles of opening the windows and doors. The basic setup of all simulation cases is summarized in Table~\ref{tab:cases}. In Runs 1-8, windows and doors are kept closed, where different mesh resolution and HVAC rates are applied to investigate the numerical uncertainty and the role of ventilation rates. The influence of the infected passenger's location is  considered by placing the passenger standing in the front or standing in the middle of the bus. In Runs 9-12 all windows are kept open and the bus runs at a constant speed of 25 mph (40.23 km/hr). Run 13 is designed to examine the effects of opening doors at bus stops.

\begin{table}[!ht]
\centering
\caption{Case setup including the location of the infected passenger, number and location of susceptible passsengers, grid resolution and the ventilation rate represented by the percentage of the maximum HVAC flow rate.}
\begin{ruledtabular}
\begin{tabular}{ccccc}
\bf Run $\#$ &\bf Infected passenger &\bf Susceptible passengers &\bf Grid &\bf HVAC rate \\\hline
1   &  standing front &  one standing rear  &  coarse  & maximum \\
2   &  -  &  - &  medium  &  -  \\
3   &  -  &  -  &  fine  &   \\
4   &  -  &  -  &  medium  &  50$\%$  \\
5   &  -  &  -  &  -  &  10$\%$  \\
6   &  standing middle &  -  &  -  &  maximum  \\
7   &  - &  -  &  -  &  50$\%$  \\
8   &  -  &  -  &  -  &  10$\%$  \\\hline
9   &  standing front  &  -  &  coarse  & maximum \\
10  &  -  &  -  &  medium  &  -  \\
11  &  -  &  -  &  fine  & - \\
12  &  standing middle  &  -  &  medium  & - \\\hline
13  &  standing front   &  -  &  -  & - \\
\end{tabular}%
\end{ruledtabular}
\label{tab:cases}%
\end{table}

\section{Results and discussion}

To visualize the flow field and spatial distribution of aerosol concentration, the results are displayed on two selected planes within the bus: a vertical plane at the centerline of the bus and a horizontal plane at the elevation of the mouths of the passengers standing on the front platform or sitting on rear seats.

The number of inhaled particles is adopted as a cumulative point of view to define the risk of each person as~\cite{vuorinen2020modelling}
\begin{equation}
N_{b}(t) = \int_0^{t}C\dot{V}_{b}~d\tau
\end{equation}
where $\dot{V}_{b}$ is the human breathing rate, and we assume the average rate as $\dot{V}_{b} = 0.33$ dm$^{3}$s$^{-1}$. 
While it is still unclear how much dose of virus is needed for someone to be infected~\cite{ledford2020does}, we use a conservative assumption of $N_{\rm b,~crit} = 50$. This value is based on the work of Kolinski and Schneider~\cite{kolinski2020superspreading}, where they analyzed twenty reported superspreading events during the ongoing pandemic.

\subsection{Refinement study}

The computational mesh is comprised of finite-volumes that are dominantly hexagonal.  Local refinement is used around fine features, namely the HVAC supply vents, the manikins, and parts of the bus surface such as seats, windows, and handrails. 

A grid refinement study is performed to assess the sensitivity of the results on the numerical discretization.  Three grids with resolution of 250~mm, 125~mm and 62.5~mm in the bulk of the flow domain are adopted as the coarse, medium and fine grids, with a total number of cells of 2.04, 5.87 and 11.65~million, respectively. The supply vents and mouths of people have the smallest cell sizes of 2~mm and 4~mm, which are the same for all three grids.

\begin{figure}[!ht]
    \centering
        \includegraphics[width=0.95\textwidth]{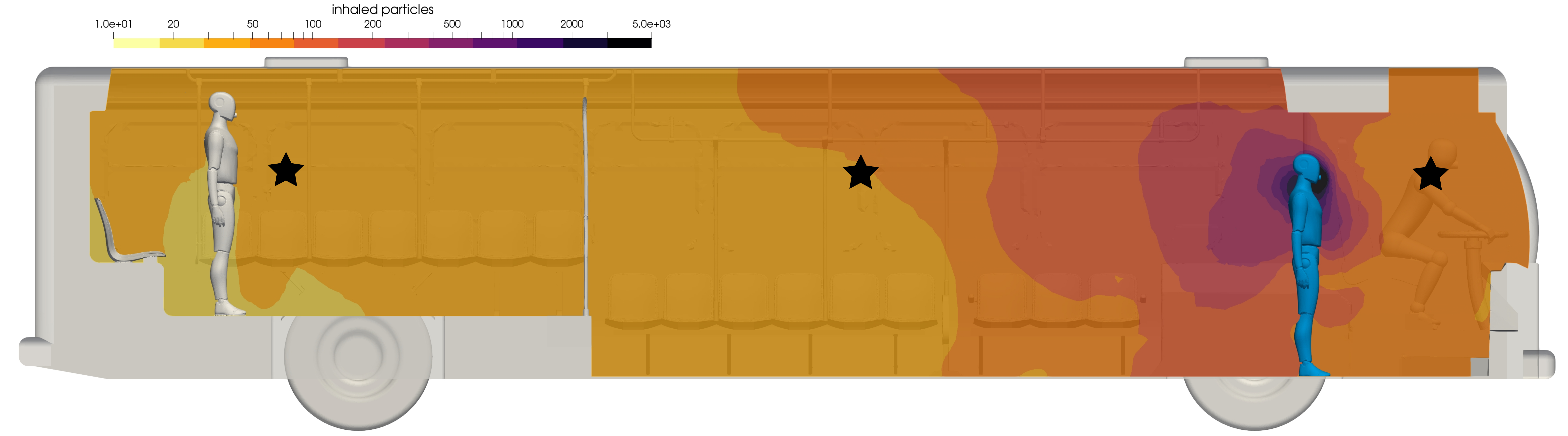}
        \caption{Contour of inhaled particles on center plane of the bus, $t = 15$ min. Black stars indicate the probe locations.}
    \label{fig:vertical}
\end{figure}

The number of inhaled particles after 15 minutes for each location on the bus for the fine grid Run~3, is shown in Fig.~\ref{fig:vertical}.  Note that the contours are spaced with logarithmic scaling. The infected passenger is placed at the front of the bus. The darker colors near the front passenger indicate that if another passenger was positioned there, the number of particles inhaled in 15~min would be much greater than the assumed threshold of 50.  In the back of the bus, the number of inhaled particles is less.  The primary mechanisms that set the distribution of particles through the bus are convection due to the air currents of the HVAC system,  the mixing due to turbulence, and the dilution due to the addition of fresh air in the HVAC system.

\begin{figure}[!ht]
    \centering
    \begin{subfigure}{0.32\textwidth}
        \includegraphics[width=\linewidth]{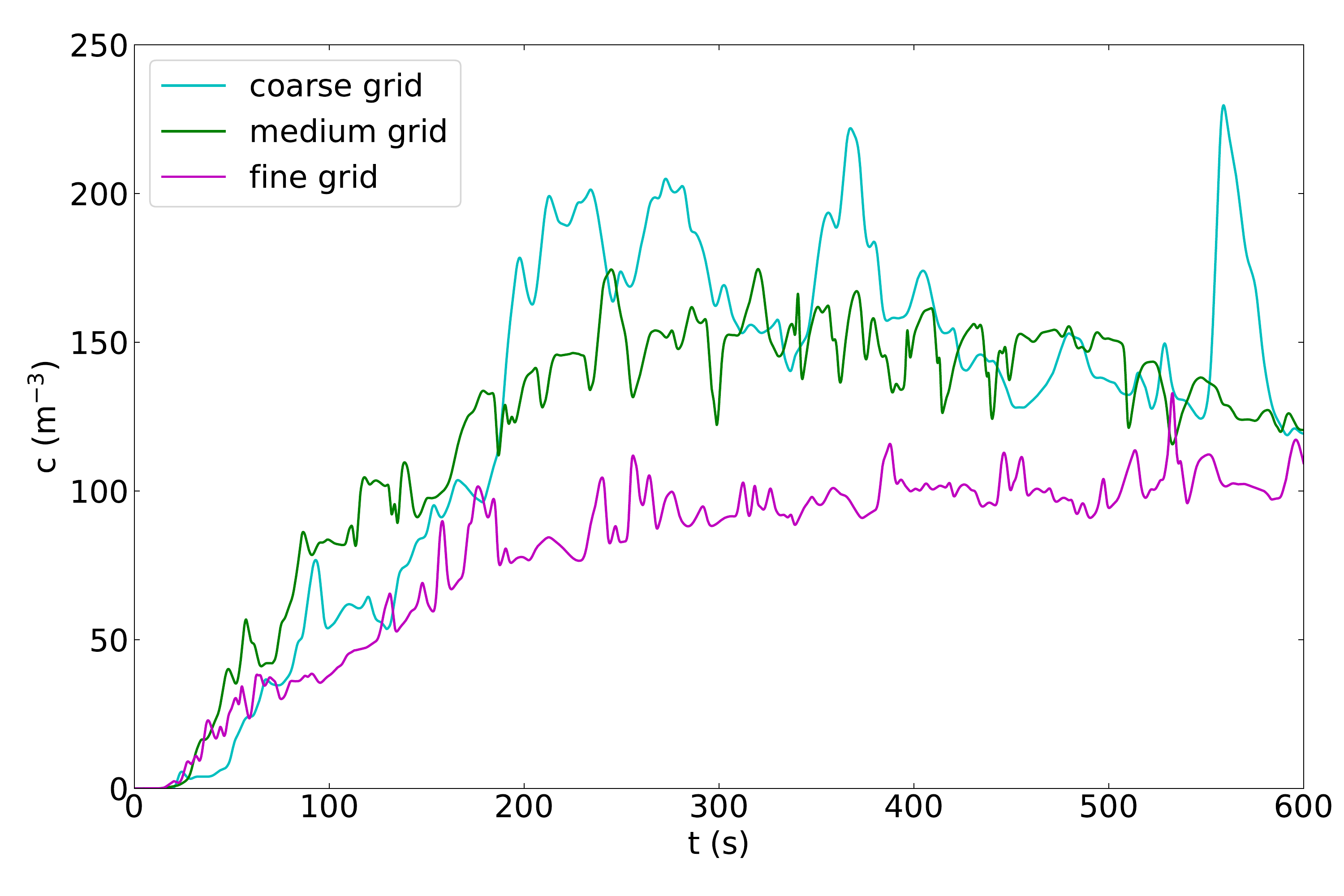}
        \caption{front probe} \label{fig:probe-front}
    \end{subfigure}
    \hspace*{\fill}
    \begin{subfigure}{0.32\textwidth}
        \includegraphics[width=\linewidth]{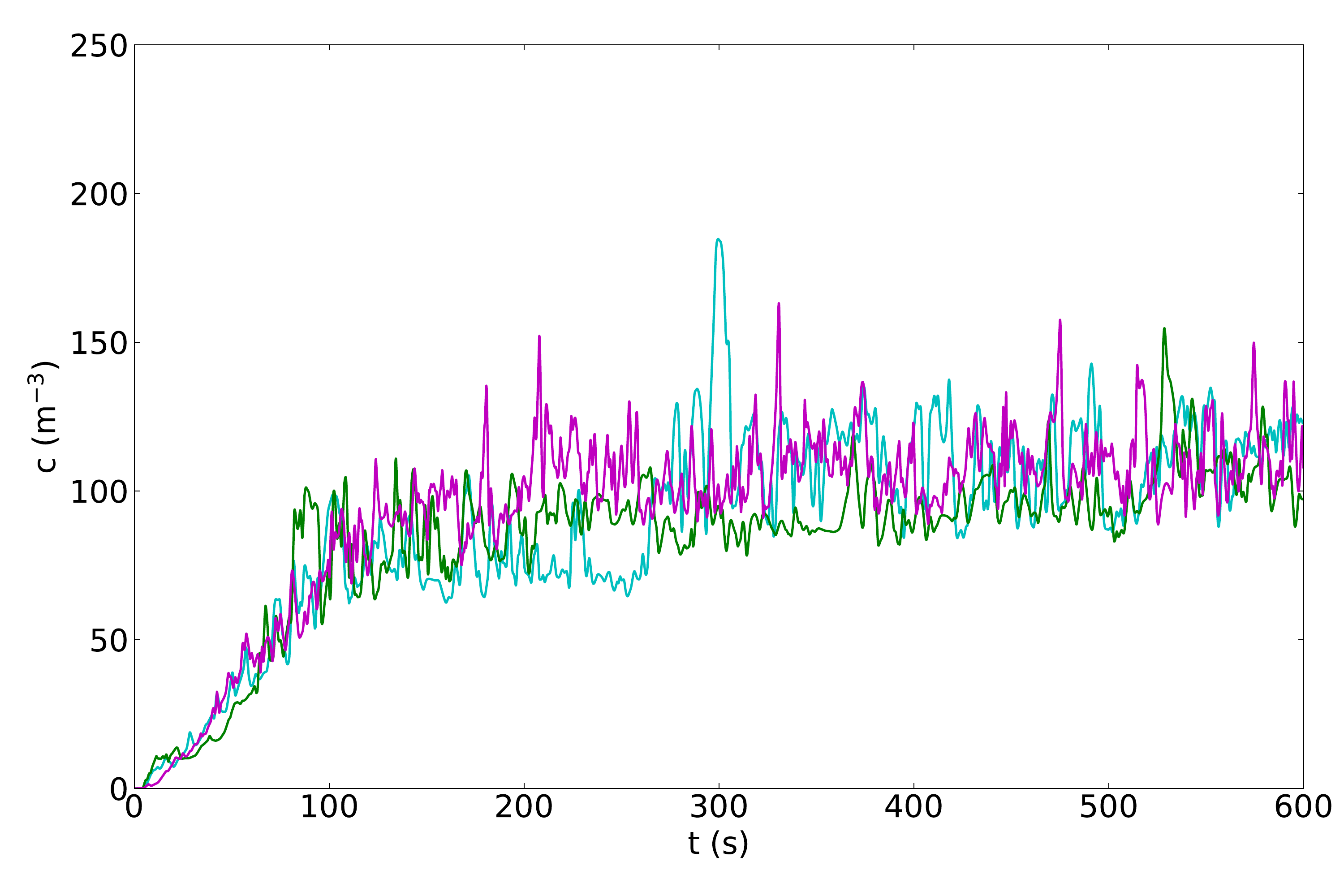}
        \caption{middle probe} \label{fig:probe-middle}
    \end{subfigure}
    \hspace*{\fill}
    \begin{subfigure}{0.32\textwidth}
        \includegraphics[width=\linewidth]{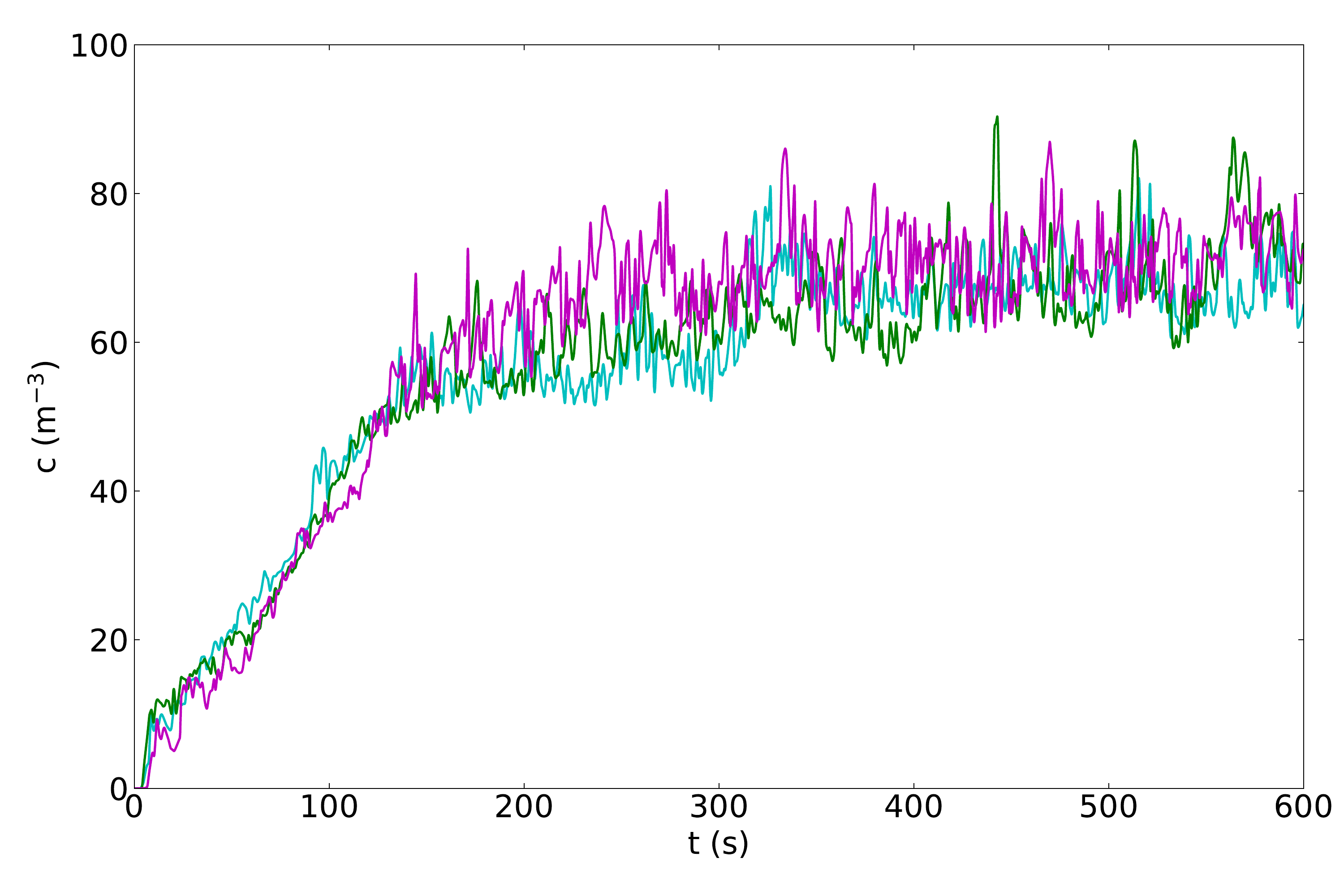}
        \caption{rear probe} \label{fig:probe-rear}
    \end{subfigure}
    \caption{Time histories of concentration with different grid resolution.}
    \label{fig:refinement}
\end{figure}

The time histories of concentration at three locations in the bus are shown in Fig.~\ref{fig:refinement}. The locations are shown as black stars in Fig.~\ref{fig:vertical}.
Inspection of the time histories shows that the three grids predict the same concentration field at all three locations, although the small differences are greatest for the probe in the front of the bus.  Also, the time history shows that the equilibrium concentration is reached in approximately 100~s for the middle probe, and around 150~s for the rear probe.  This indicates that even a short trip on a bus can present exposure to a passenger, although the quantity of inhaled particles will be small at first, and grow with time. Based on the analysis of Runs~1-3 the medium grid is used for the rest of the Runs for the windows-closed simulations.

\subsection{Flow-field inside bus}

\begin{figure}[!ht]
    \centering
    \begin{subfigure}{1\textwidth}
        \includegraphics[width=0.95\textwidth]{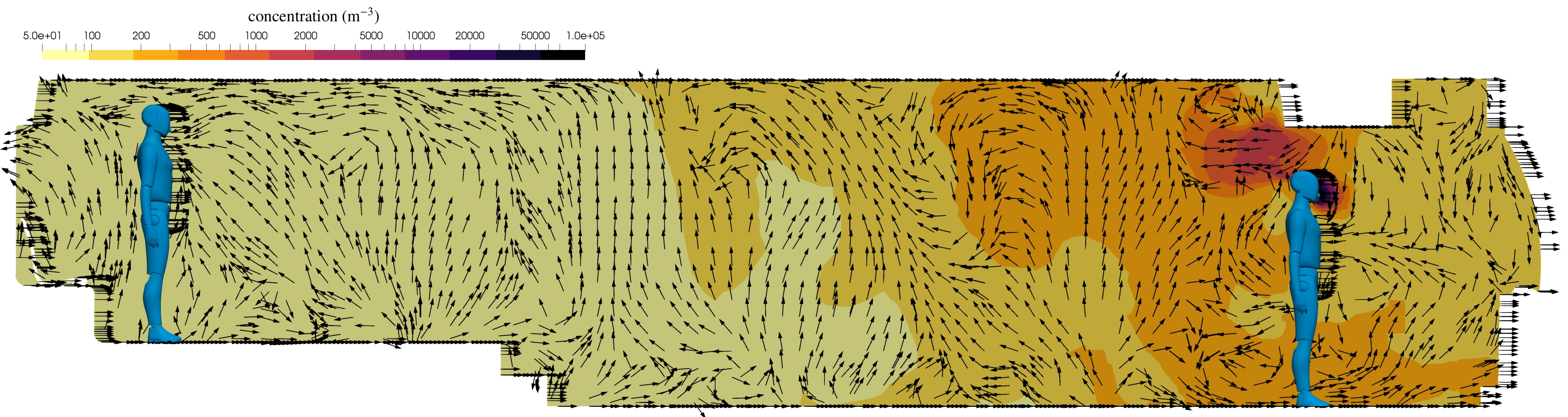}
        \caption{Contour of aerosol concentration and velocity vectors, $t = 15$ min.}
        \end{subfigure}
            \begin{subfigure}{1\textwidth}
        \includegraphics[width=0.95\textwidth]{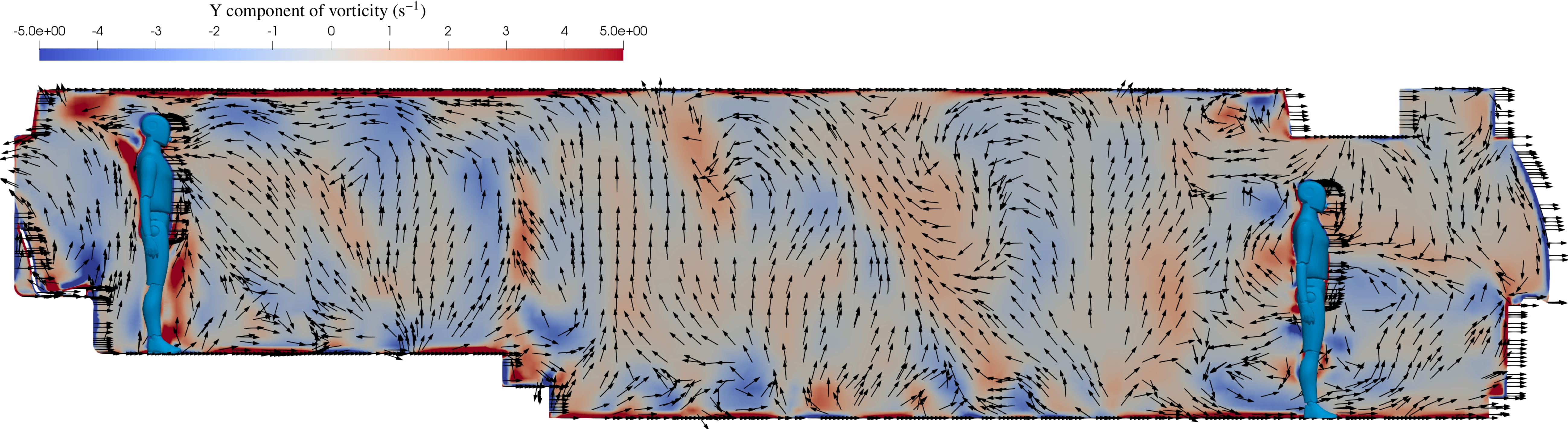}
        \caption{Contour of vorticity and velocity vectors, $t = 15$ min.}
        \end{subfigure}
        \caption{Flow field details on center plane of the bus with windows closed and HVAC at maximum rate.}
    \label{fig:verticalConcentrationVelocity}
\end{figure}

Turbulence is a primary transport mechanism for  aerosols, and it depends on the geometry of the passenger compartment, the opening of doors and windows, and the HVAC system.  The high-resolution simulations used in this work allow for inspection of the dominant flow features in the passenger compartment.  Fig.~\ref{fig:verticalConcentrationVelocity} shows the velocity vector field on the center plane together with the concentration field (top) and the vorticity field (bottom). It can be seen that the turbulent flow moves both up and down as the net flow is rearward through the compartment. The highest values of vorticity are observed near corners and the supply vents, which aid in mixing aerosol concentration.

\subsection{Risk under different ventilation rates}

The HVAC system is a primary aspect of the transport of aerosols within the bus (this is true for many confined spaces), and it is important to quantitatively assess how the exposure varies as the HVAC fan speed is changed.  The HVAC system adds fresh air as a fraction of its flow rate (in this case 20\%), and it acts to mix and transport the smallest particles through the cabin.    

\begin{figure}[!ht]
    \captionsetup[subfigure]{slc=off,margin={0.1cm,0cm}}
    \centering
    \begin{subfigure}{0.95\textwidth}
        \caption{maximum HVAC rate}
        \includegraphics[width=\linewidth]{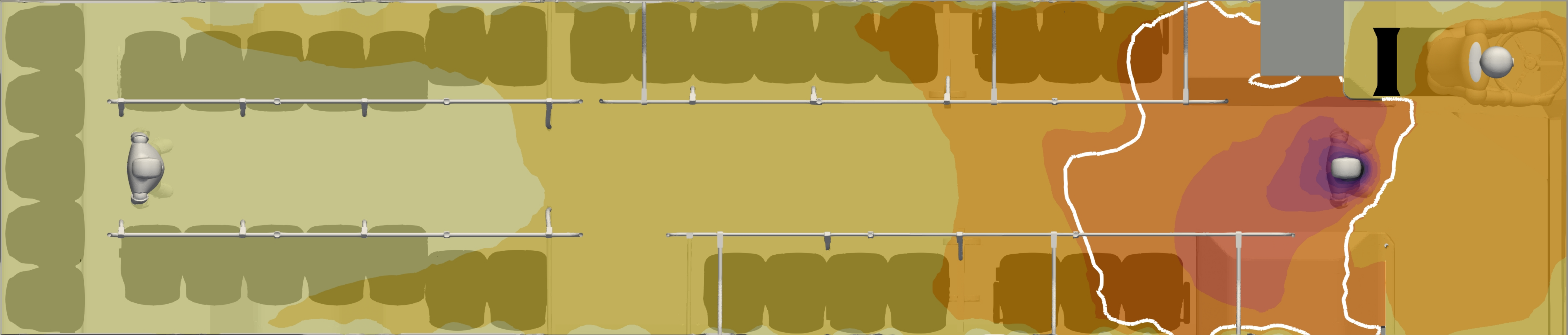} 
    \end{subfigure}
    \medskip
    \vspace{0.2cm}
    \begin{subfigure}{0.95\textwidth}
        \caption{50$\%$ of maximum rate}
        \includegraphics[width=\linewidth]{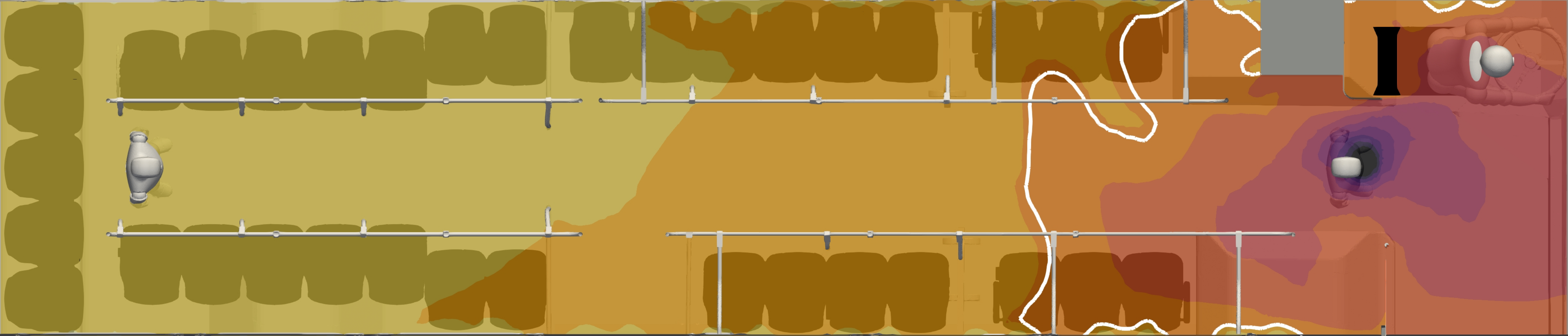}   
    \end{subfigure}
    \medskip
    \vspace{0.2cm}
    \begin{subfigure}{0.95\textwidth}
        \caption{10$\%$ of maximum rate}
        \includegraphics[width=\linewidth]{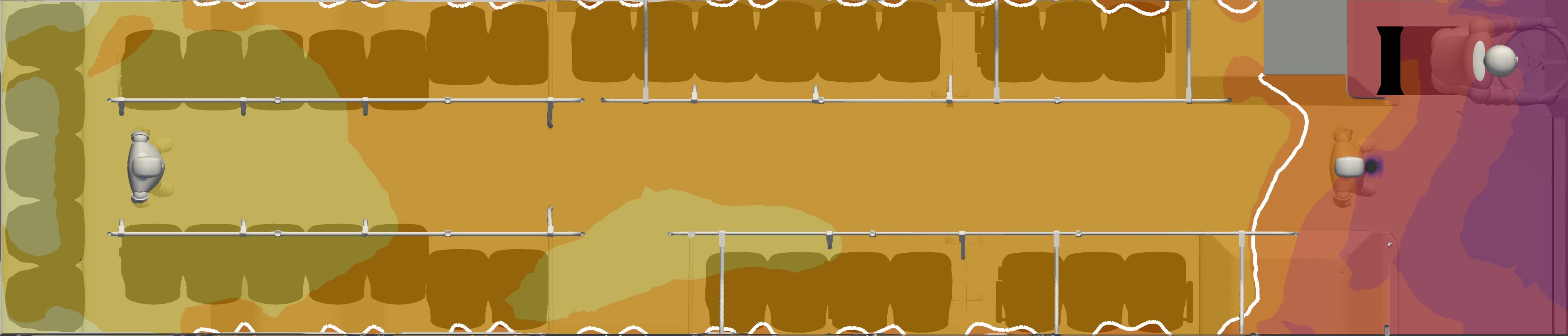}   
    \end{subfigure}
    \medskip
    \vspace{0.2cm}
    \begin{subfigure}{0.7\textwidth}
        \includegraphics[width=\linewidth]{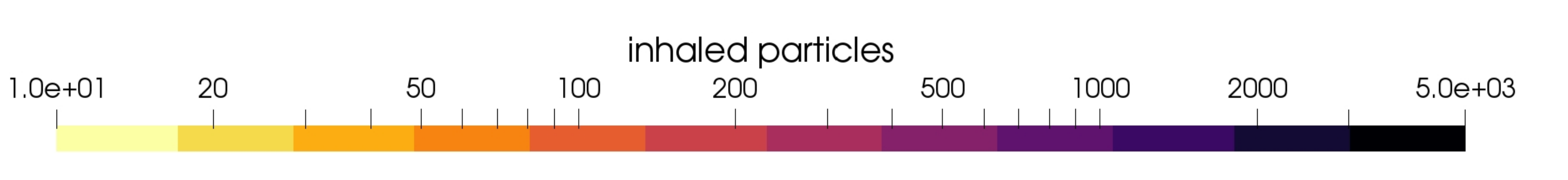}   
    \end{subfigure}
    \caption{Contours of inhaled particles for different HVAC rates with the infected passenger standing in the front of the bus at t~=~15 min. The white contour lines represent the critical number of inhaled particles $N_{\rm b,~crit} = 50$.}
    \label{fig:HVAC-front}
\end{figure}

\begin{figure}[!ht]
    \captionsetup[subfigure]{slc=off,margin={0.1cm,0cm}}
    \centering
    \begin{subfigure}{0.95\textwidth}
        \caption{maximum HVAC rate}
        \includegraphics[width=\linewidth]{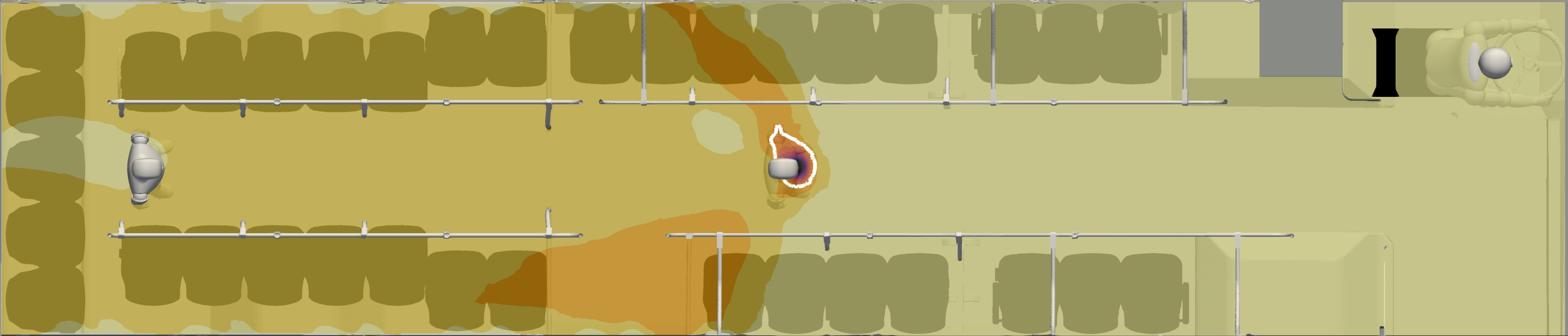} 
    \end{subfigure}
    \medskip
    \vspace{0.2cm}
    \begin{subfigure}{0.95\textwidth}
        \caption{50$\%$ of maximum rate}
        \includegraphics[width=\linewidth]{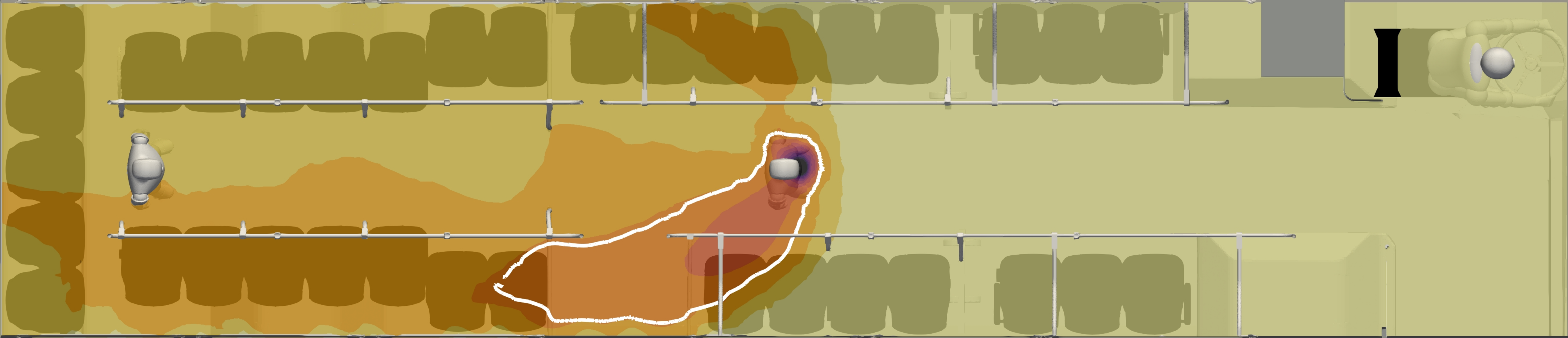}   
    \end{subfigure}
    \medskip
    \vspace{0.2cm}
    \begin{subfigure}{0.95\textwidth}
        \caption{10$\%$ of maximum rate}
        \includegraphics[width=\linewidth]{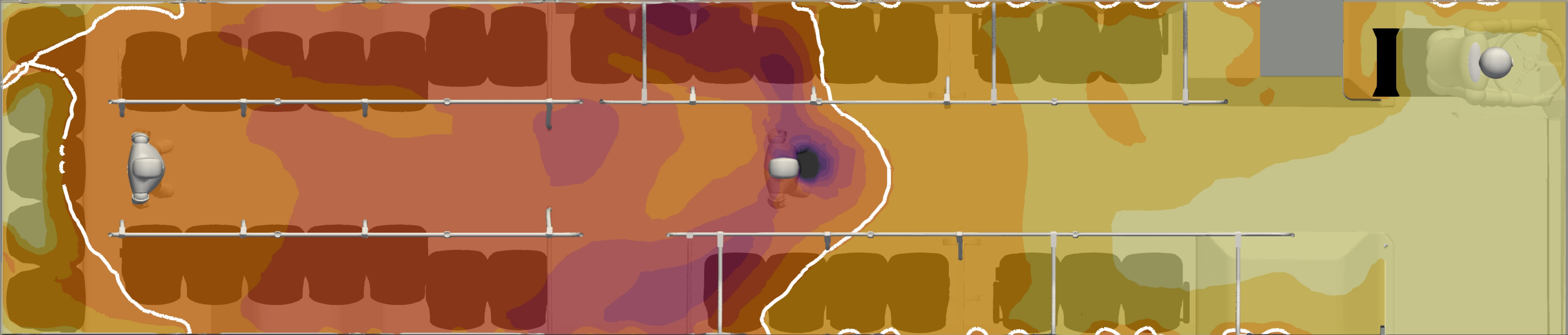}   
    \end{subfigure}
    \medskip
    \vspace{0.2cm}
    \begin{subfigure}{0.7\textwidth}
        \includegraphics[width=\linewidth]{./fig/legend.pdf}   
    \end{subfigure}
    \caption{Contours of inhaled particles for different HVAC rates with the infected passenger standing in the middle of the bus at t~=~15 min. The white contour lines represent the critical number of inhaled particles $N_{\rm b,~crit} = 50$.}
    \label{fig:HVAC-middle}
\end{figure}

Figs.~\ref{fig:HVAC-front} and~\ref{fig:HVAC-middle} shows the contours of inhaled particles for three different HVAC settings: the maximum flow rate, 50\% of the maximum, and 10\% of the maximum.  In Fig.~\ref{fig:HVAC-front} the infected passenger is at the front of the bus, and in Fig.~\ref{fig:HVAC-middle} the infected passenger is in the middle. The contour representing the inhalation of 50 particles is shown in the thick white line such that inside this contours a passenger would inhale more than 50, and outside they would inhale less.  Hence a count of the number of seats or standing positions inside the area bounded by the white line indicates the number of transmissions in the 15~min exposure time.

In Fig.~\ref{fig:HVAC-front} it can be seen that as soon as the HVAC rate is reduced to 50\%, the region of elevated risk grows substantially, and covers the entire front of the bus.  The case with the infected passenger in front poses serious risk for the driver especially considering that the driver is on the bus for extended periods of time.

Fig.~\ref{fig:HVAC-middle} shows the influence of the HVAC flow rate for the infected passenger in the middle of the bus.  A similar effect is seen where a reduction in the fan speed enhances risk to surrounding passengers, and while the area of greater than 50 particles is relatively small for the 50\% fan speed, for the lowest fan speed nearly all passengers to the rear of the infected passenger could be infected during the 15~min trip.  In this case, the driver is relatively safe since the single HVAC return vent draws air towards the rear of the bus, and effectively isolates the driver.  Note that for the lowest flow rate, while the transport of the aerosols is primarily rearward, there is transport forward of the infected passenger.  This is due to the chaotic nature of the turbulent flow, as well as diffusion which becomes more important as the ambient air currents lessen in intensity. Also, while the HVAC adds 20\% fresh air, it does take the virus-laden air and returns it throughout the HVAC supply vents that are located from front to rear.

To summarize the numbers of transmissions in the 15-minute exposure, when the infected passenger is at the front of the bus, there will be 3, 5 and 1 transmissions for the 100\%, 50\% and 10\% HVAC rates, respectively. The numbers are 0, 2 and 23 when the infected passenger stands in the middle of the bus.

\subsection{Effects of face masks}

Face masks (or face coverings) are a primary line of defense for reducing COVID-19 transmission.  Many researchers around the world are working to scientifically quantify the effects of wearing masks.  In this work, we use a simple model for a mask based on the work of~\cite{Dbouk20}, in which the fraction of exhaled particles is predicted using CFD.  Face masks are found in different types, and in this work two masks, a surgical mask and a handmade mask are analyzed.  We assume the surgical mask will block 90\% of the exhaled and inhaled aerosols, and the handmade masks block 30\% of the particles.

\begin{figure}[!ht]
    \captionsetup[subfigure]{slc=off,margin={0.1cm,0cm}}
    \centering
    \begin{subfigure}{0.95\textwidth}
        \caption{nobody wears a mask} \label{fig:run8}
        \includegraphics[width=\linewidth]{./fig/run8.pdf} 
    \end{subfigure}
    \medskip
    \vspace{0.2cm}
    \begin{subfigure}{0.95\textwidth}
        \caption{everyone wears a surgical mask} \label{fig:run8-Both-Surgical}
        \includegraphics[width=\linewidth]{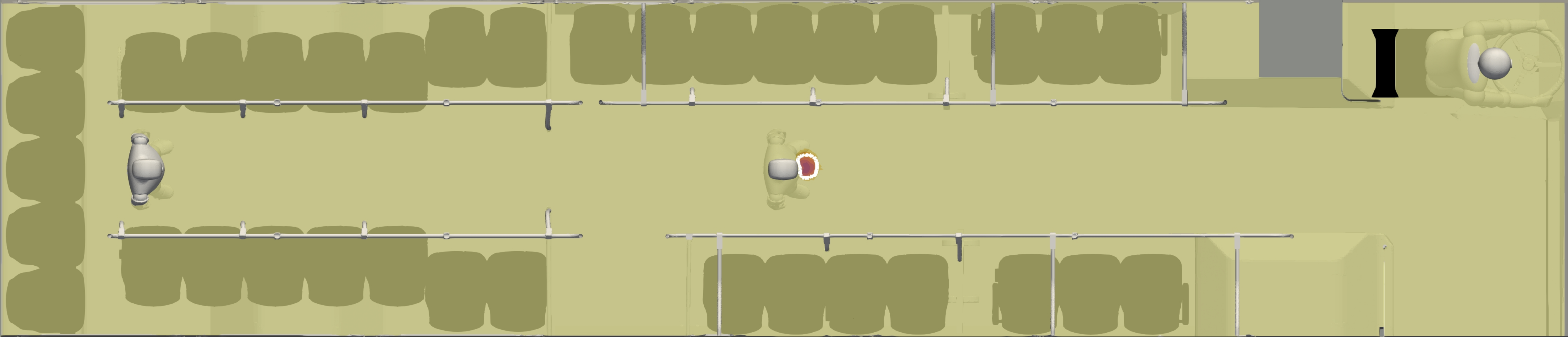}   
    \end{subfigure}
    \medskip
    \vspace{0.2cm}
    \begin{subfigure}{0.95\textwidth}
        \caption{everyone wears a handmade mask} \label{fig:run8-Both-DIY}
        \includegraphics[width=\linewidth]{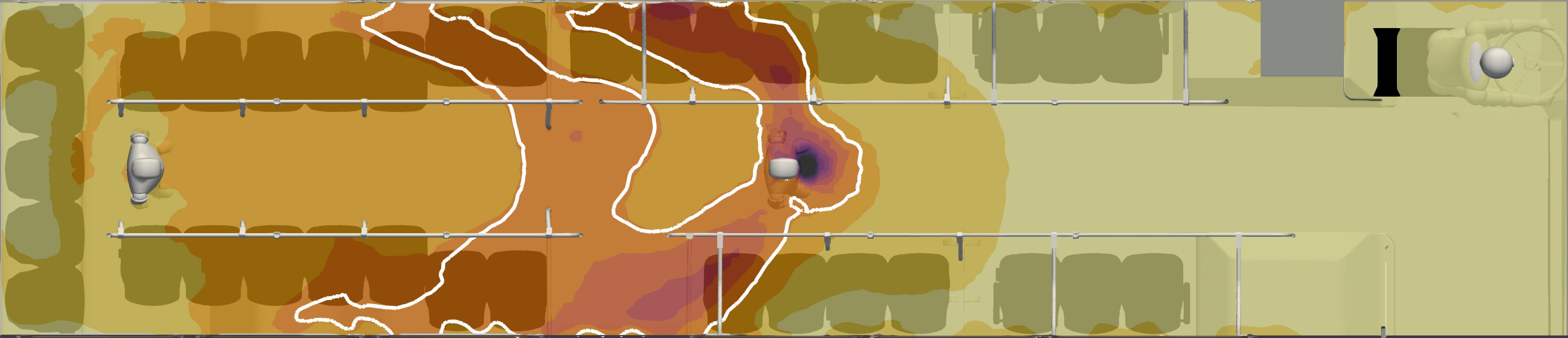}   
    \end{subfigure}
    \medskip
    \vspace{0.2cm}
    \begin{subfigure}{0.7\textwidth}
        \includegraphics[width=\linewidth]{./fig/legend.pdf}   
    \end{subfigure}
    \caption{Contours of inhaled particles for different scenarios of face coverings at t~=~15 min. The white contour lines represent the critical number of inhaled particles $N_{\rm b,~crit} = 50$.}
    \label{fig:mask}
\end{figure}

Fig.~\ref{fig:mask} shows the contours of inhaled particles for the cases of no mask (top), everyone with a surgical mask (middle), and everyone with a handmade mask (bottom).  It is impressive to see how the surgical mask significantly reduces the number of inhaled particles. In the top figure with no mask, nearly all passengers to the rear of the infected passenger inhale more than 50. On the other hand, when everyone wears a surgical mask, during the 15 minute ride, not a single passenger inhales anywhere near 50 particles, and unless the susceptible person is standing face-to-face with the infected person, the number of inhaled particles is less than two.

In the bottom of Fig.~\ref{fig:mask} the number of particles for a handmade mask is shown.  Clearly the effect of wearing a mask is to reduce the number of aerosols that are inhaled, although in this case there are still several people that could be inside the white contour. Also, for passengers throughout the rear portion of the bus, the effect of handmade masks is to reduce the number of particles from more than 50 in case without masks, to a number as low as 20. For this case, 23 seated passengers will be infected during the 15-minute ride if no one wears a mask, the number will be 0 if both the infected and susceptible passengers wear surgical masks and 9 if both wear handmade masks.

\subsection{Effects of opening windows and doors}

An important mechanism for reducing the aerosol concentration is to add fresh air.  This can be done manually by adjusting the HVAC system, or passively when the doors and windows are open on the bus.  To quantify the effect of opening windows and doors, simulations are conducted with the bus moving at 25~mph (40.23 km/hr) with the windows open.  Also a simulation is done with the windows closed but with the doors open for 30~s at five stops during the 15 min trip.  For the case with doors open, there is a 5 mph (8.05 km/hr) wind blowing opposite of the direction of travel of the bus.

\begin{figure}[!ht]
    \centering
    \begin{subfigure}{1\textwidth}
        \includegraphics[width=0.95\textwidth]{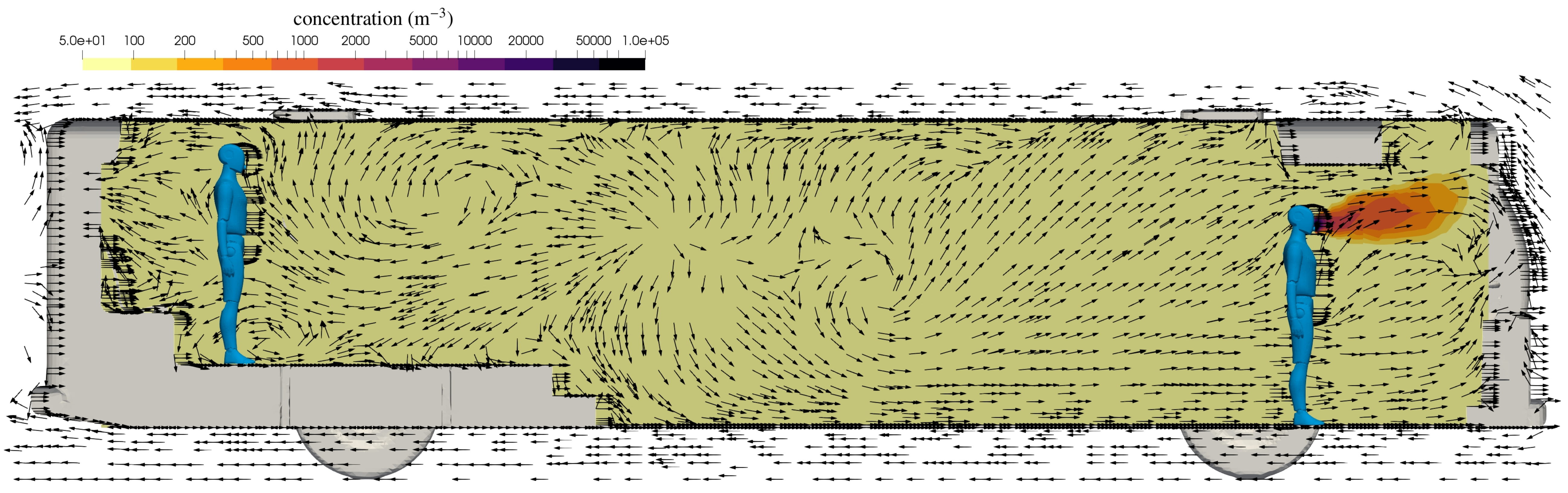}
        \caption{Contour of aerosol concentration and velocity vectors, $t = 15$ min.}
        \end{subfigure}
            \begin{subfigure}{1\textwidth}
        \includegraphics[width=0.95\textwidth]{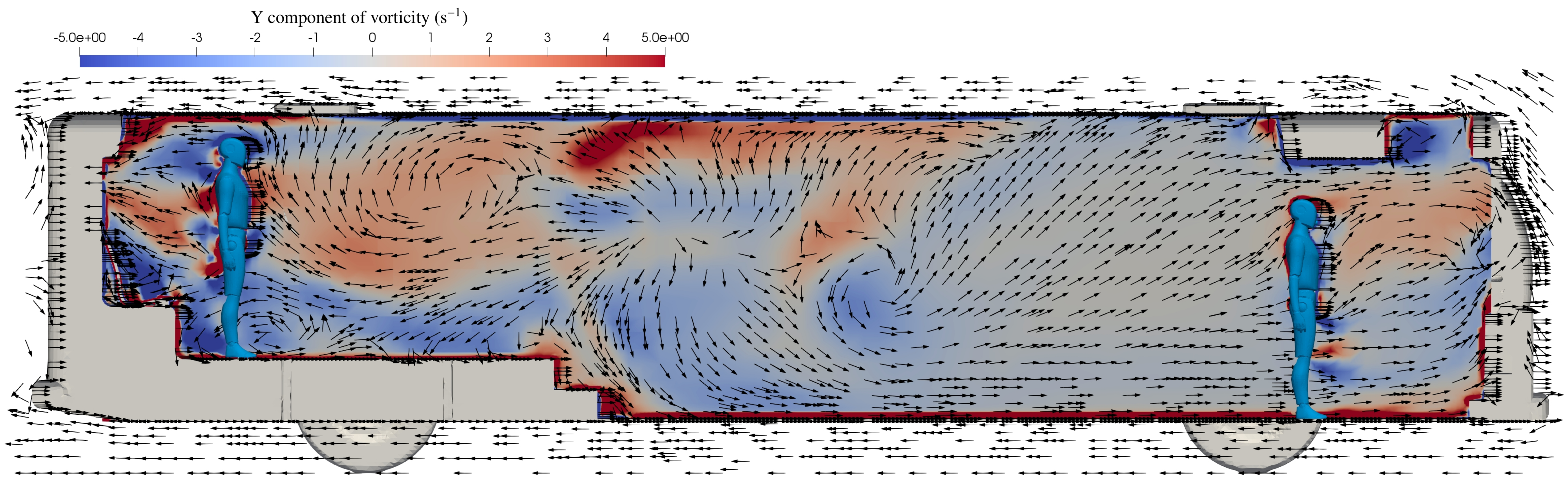}
        \caption{Contour of vorticity and velocity vectors, $t = 15$ min.}
        \end{subfigure}
        \caption{Flow field details on center plane of the bus with windows open.}
    \label{fig:verticalConcentrationVelocityOpen}
\end{figure}

The primary difference when the doors and windows open is that fresh air can be added (or virus-laden air removed), and the flow field can be materially different.  Fig.~\ref{fig:verticalConcentrationVelocityOpen} shows the flow field inside the bus with the windows open. This figure should be compared to Fig.~\ref{fig:verticalConcentrationVelocity} where the windows are closed. The most notable difference is that when the windows are open, there is a net rearward flow for the middle of the bus and behind, but there is a net outflow through the driver window which draws air forward. This highlights the complicated nature of turbulent flows within occupied spaces, and how small changes can significantly alter the flow and hence risk of transmission.  While in aggregate the risk is reduced for passengers when the windows are opened, the risk to the driver has been increased when the infected passenger is standing up front in the bus.  Similarly to the windows closed flow field the movement of aerosols has both upward and downward motion that mixes and renders risk the same whether one is seated or standing.

\begin{figure}[!ht]
    \captionsetup[subfigure]{slc=off,margin={0.1cm,0cm}}
    \centering
    \begin{subfigure}{0.95\textwidth}
        \caption{bus is enclosed} \label{fig:run2}
        \includegraphics[width=\linewidth]{./fig/run2.pdf} 
    \end{subfigure}
    \medskip
    \vspace{0.2cm}
    \begin{subfigure}{0.95\textwidth}
        \caption{windows are open} \label{fig:wind-run2}
        \includegraphics[width=\linewidth]{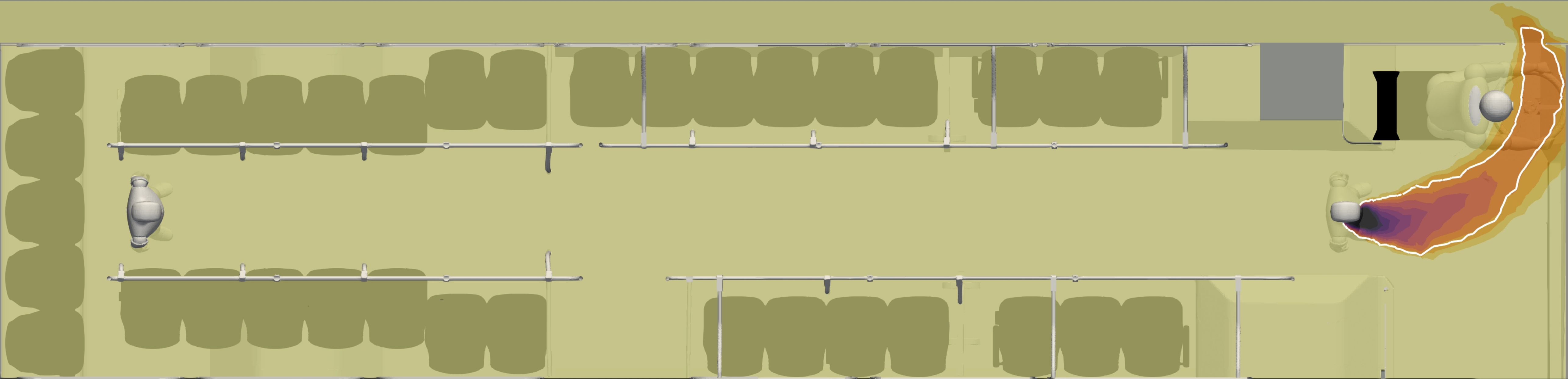}   
    \end{subfigure}
    \medskip
    \vspace{0.2cm}
    \begin{subfigure}{0.95\textwidth}
        \caption{doors are open at each stop} \label{fig:door}
        \includegraphics[width=\linewidth]{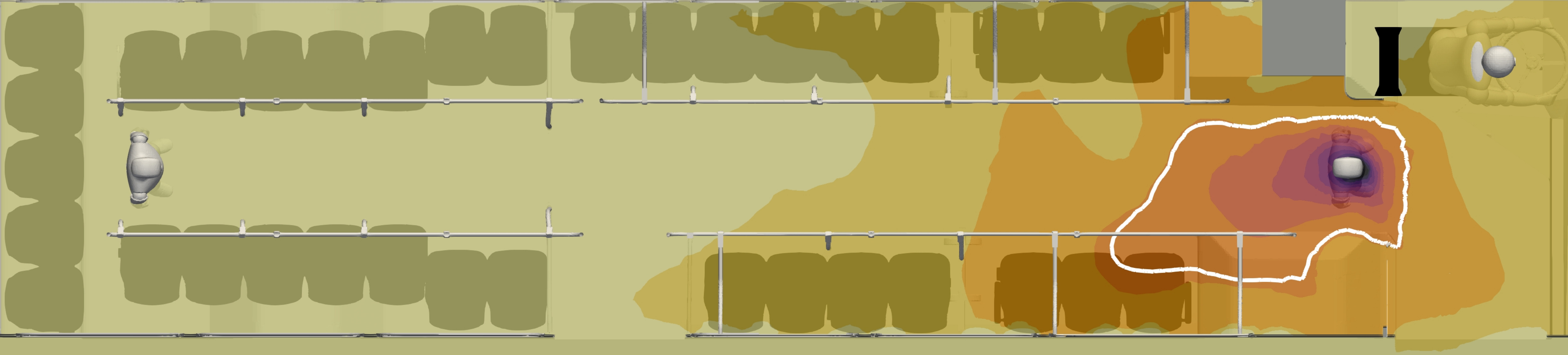}   
    \end{subfigure}
    \medskip
    \vspace{0.2cm}
    \begin{subfigure}{0.7\textwidth}
        \includegraphics[width=\linewidth]{./fig/legend.pdf}   
    \end{subfigure}
    \caption{Contours of inhaled particles with different setup for the windows and doors at t~=~15 min. The white contour lines represent the critical number of inhaled particles $N_{\rm b,~crit} = 50$.}
    \label{fig:window-door}
\end{figure}

Fig.~\ref{fig:window-door} summarizes the results for the effects of opening windows and doors. At the top of this figure the result for a single passenger without facemask is shown, in the middle the contour of inhaled particles for the windows open, and the bottom for the case when the windows are closed yet the doors open periodically.  Here is clearly seen how overall the number of inhaled particles decreases significantly throughout the passenger compartment, with the exception of focusing of virus-laden air in front of the driver.

The influence of opening the doors is seen to slightly reduce the number of particles inhaled throughout the bus.

To summarize the numbers of transmissions in the 15-minute exposure, there will be 3 transmissions in the enclosed cabin, 1 transmission (the driver) if windows are open and 1 transmission if doors open periodically.

\section{Conclusions}

In this paper, a detailed analysis of the airborne transmission of respiratory aerosols is conducted using experimental measurement and computational fluid dynamics.  The transmission on an urban bus is studied to identify the transmission mechanisms and to assess strategies to reduce risk.   Specifically, risk is quantified and the effects of the air-conditioning system, opening windows and doors, and wearing masks are analyzed.

Experiments are performed on a University of Michigan campus bus. The temporal history of concentration and size of particles emitted from a smoke generator are measured for a variety of particle injection and sampling locations throughout the bus.  The effects of opening doors and windows are quantified.

Numerical simulations are performed with a highly infectious passenger aboard the bus, and the exhaled aerosols are modeled as a  concentration field.  The transport of the aerosol concentration is determined by the solution of the turbulent flow within the passenger compartment and a transport equation for the concentration.  A risk metric of the number of particles inhaled by susceptible passengers is defined so that different risk mitigation strategies can be compared and assessed quantitatively.

The analysis shows that under the condition of the HVAC system at its maximum setting the airflow in the bus is turbulent and the time scales of transit from an infected passenger to any susceptible passenger is less than a minute.  In other words, six-foot spacing does not protect a susceptible passenger.  While the short response time appears to increase risk, the HVAC actually reduces risk because the turbulence mixes the aerosols with the ambient air thereby reducing concentration, and the HVAC system adds fresh air that thus dilutes the concentration further.

The effect of opening doors and windows is to reduce the concentration by approximately one half.  The CFD analysis shows that for almost all passengers this is true, while care should be exercised that in certain cases the outflow of contaminated cabin air could pass by a passenger (in this case it is the driver), and increase the risk to those near the outflow window or door.

A mask model is used to quantify and visualize their influence. It is shown that well fitted surgical masks, when worn by both infected and susceptible passengers, can nearly eliminate the transmission of the disease.  In the case of poorer quality mask their effect is still to reduce transmission for all aboard the bus.

\begin{acknowledgments}
The computing resources provided by the staff of the Advanced Research Computing at the University of Michigan, Ann Arbor are greatly appreciated.
\end{acknowledgments}

\section{Data Availability}

The data that support the findings of this study are available from the corresponding author upon reasonable request.

\nocite{*}
\bibliography{myRef.bib}

\end{document}